\documentclass[notitlepage,a4paper,aps,prd,onecolumn,superscriptaddress,nofootinbib,groupedaddress,longbibliography]{revtex4-2}
\usepackage[utf8]{inputenc}
\usepackage[T1]{fontenc}
\usepackage[svgnames]{xcolor}
\usepackage{amsfonts}
\usepackage{amssymb}
\usepackage{amsmath}
\usepackage[colorlinks,hyperindex,plainpages=false]{hyperref}
\usepackage[final,inline]{showlabels}

\newcommand{\dd}{\mathrm{d}}
\DeclareMathOperator{\artanh}{artanh}
\DeclareMathOperator{\sgn}{sgn}

\begin{document}

\title{Bianchi cosmologies with four-dimensional groups of motion in New General Relativity}

\author{Manuel Hohmann}
\email{manuel.hohmann@uni-oldenburg.de}
\affiliation{Laboratory of Theoretical Physics, Institute of Physics, University of Tartu, W. Ostwaldi 1, 50411 Tartu, Estonia}

\author{Paul Martin Kull}
\email{paul.martin.kull@ut.ee}
\affiliation{Laboratory of Theoretical Physics, Institute of Physics, University of Tartu, W. Ostwaldi 1, 50411 Tartu, Estonia}

\author{Margus Saal}
\email{margus.saal@ut.ee}
\affiliation{Laboratory of Theoretical Physics, Institute of Physics, University of Tartu, W. Ostwaldi 1, 50411 Tartu, Estonia}

\begin{abstract}
We study the dynamics of the most general metric teleparallel spacetimes of Bianchi types II, III and IX, which admit a four-dimensional symmetry group, in the 1-parameter family of New General Relativity. We first show that in each of these classes the most general metric teleparallel geometry is described by four free functions of time, while another four functions of time describe the most general energy-momentum tensor compatible with the same symmetries. We then derive the field equations of New General Relativity for this class of geometries. Finally, we present some solutions to these field equations. Our results show that both solutions resembling features of the standard Friedmann-Lemaître-Robertson-Walker cosmology of General Relativity and deviations thereof exist.
\end{abstract}

\maketitle


\section{Motivation}\label{sec:motivation}
Recent observations in cosmology, such as various measurements of the present-time value \(H_0\) of the Hubble parameter by either observations of the late-time evolution of the universe or inference from early-time observations, put tensions on the so-called $\Lambda$CDM model~\cite{Planck:2018vyg,DiValentino:2021izs}. The latter models the dynamics of the universe through general relativity, a cosmological constant $\Lambda$ and cold dark matter (CDM), and has emerged as the standard model of cosmology. Numerous potential explanations for these seemingly contradictory observations have been offered, including new and unknown types of matter, or modifications of the underlying theory of gravity, and their effects on cosmological observations have been studied~\cite{Heisenberg:2018vsk,CANTATA:2021asi,CosmoVerseNetwork:2025alb}.

Among the modifications in the gravity sector, there is a class of theories in which gravity is not mediated by the curvature of the Levi-Civita connection, but by the torsion of a flat, metric-compatible connection, known as (metric) teleparallel gravity~\cite{Aldrovandi:2013wha,Golovnev:2018red,Hohmann:2022mlc}. For the cosmological setting, the most general homogeneous and isotropic backgrounds have been derived~\cite{Hohmann:2020zre,Coley:2022qug} and the resulting cosmology has been extensively studied~\cite{Bahamonde:2021gfp,Cai:2015emx}. New General Relativity (NGR)~\cite{Hayashi:1979qx} is a specific teleparallel theory constructed from the three possible quadratic torsion scalars with arbitrary coefficients, and its 1-parameter subclass (1PNGR) has been proven to be stable and ghost-free~\cite{Hayashi:1979qx}. However, teleparallel gravity theories, including 1PNGR, face a significant theoretical challenge known as the strong coupling problem, where certain modes in linear perturbation theory become non-dynamical for highly symmetric background solutions such as the homogeneous and isotropic FLRW cosmology~\cite{Golovnev:2018wbh,Golovnev:2020zpv} and Minkowski spacetime~\cite{BeltranJimenez:2019nns,Golovnev:2020nln,BeltranJimenez:2020fvy,Bahamonde:2024zkb}. This issue arises because the degrees of freedom of the theory do not appear at linear order in these symmetric settings~\cite{Blagojevic:2020dyq,Guzman:2019oth}. One promising approach to circumvent this problem is to consider Bianchi spacetimes, which retain the notion of symmetry but employ smaller symmetry groups than conventionally assumed in cosmology. These spatially homogeneous, anisotropic metrics have been fully classified~\cite{Bianchi:1898,Bianchi:2001}, and several of them have been studied in teleparallel gravity~\cite{Rodrigues:2012qua,Rodrigues:2014xam,Paliathanasis:2022vux,Golovnev:2025lpt,Rathore:2025cey}. Following this approach may provide backgrounds where the full degrees of freedom become dynamical, potentially curing the strong coupling problem~\cite{Barker:2022kdk,Hu:2023juh,Hu:2024qjs}.

Among the different classes of spatially homogeneous spacetimes, three classes allow for a one-dimensional isotropy group, so that their full symmetry group is four-dimensional. The most general metric teleparallel geometries obeying these symmetry groups have been found~\cite{Hohmann:2023sto}, using the general construction method for teleparallel geometries with spacetime symmetries~\cite{Hohmann:2019nat}, and it has been shown that for each class the most general geometry is fully determined by six free functions of time. This is the starting point for the work presented in this article. For each class, we first discuss the residual coordinate freedom, and show that two free functions can be absorbed by a coordinate transformation, so that only four free functions remain. We then apply the same symmetries to the energy-momentum tensor and show that the most general solution is described by an anisotropic fluid, which is determined by four further functions of time. Finally, we insert the teleparallel background geometries and corresponding energy-momentum tensors into the field equations of 1PNGR, and make use of the method of dynamical systems to conclude on particular solutions of the resulting cosmological dynamics.

The outline of this article is as follows. In section~\ref{sec:ngr}, we briefly review 1PNGR and introduce the conventions used in this article. We then review the most general teleparallel spacetimes with four-dimensional symmetry groups in section~\ref{sec:geometry} and show how two free functions can be absorbed by coordinate transformations. We complement these geometries by energy-momentum tensors obeying the same symmetries in section~\ref{sec:enmom}, and make use of the energy-momentum conservation to solve for one of its components. We make use of the results from the preceding two sections in section~\ref{sec:cosmofield} in order to derive the dynamical equation for the discussed cosmological models and write them in the form of a dynamical system. Finally, in section~\ref{sec:solution}, we discuss several solutions. We end with a conclusion in section~\ref{sec:conclusion}.

\section{New General Relativity}\label{sec:ngr}
We begin with a brief overview of the mathematical formalism of NGR. Like any teleparallel theory, the fundamental dynamical variables are the tetrad $\theta^a={\theta^a}_\mu dx^\mu$ and spin connection ${\omega^a}_b={\omega^a}_{b\mu}dx^\mu$. The connection is flat and metric-compatible, but contains torsion. The tetrad and spin connection can be used to construct the metric and torsion tensors:
\begin{gather}\label{eq:metric+torsion}
g_{\mu\nu}=\eta_{ab}{\theta^a}_\mu {\theta^b}_\nu\,,\quad {T^a}_{\mu\nu}=2\left(\partial_{[\mu}{\theta^a}_{\nu]}+{\omega^a}_{b[\mu}{\theta^b}_{\nu]}\right)\,,
\end{gather}
where $\eta_{ab}=\operatorname{diag}(-1,1,1,1)$ in our treatment. The spin connection ensures invariance under Lorentz transformations and constitutes a gauge freedom due to its flatness, which means that we can choose the so-called Weitzenböck gauge, where ${\omega^a}_{b\mu}=0$. This leaves the tetrad as the only fundamental dynamical variable. The torsion tensor admits an irreducible decomposition into three parts:
\begin{gather}\label{eq:torsion_irr}
    a_\mu=\frac{1}{6}\epsilon_{\mu\nu\sigma\rho}T^{\nu\sigma\rho}\,,\quad v_\mu={T^\nu}_{\nu\mu}\,,\quad t_{\sigma\mu\nu}=\frac{1}{2}(T_{\sigma\mu\nu}+T_{\mu\sigma\nu})+\frac{1}{6}(g_{\nu\sigma}v_\mu+g_{\nu\mu}v_\sigma)-\frac{1}{3}g_{\sigma\mu}v_\nu\,,
\end{gather}
where $\epsilon_{\mu\nu\sigma\rho}$ is the covariant Levi-Civita tensor. These are known, respectively, as the axial, vector, and tensor torsions. We can construct a parity-even scalar from each:
\begin{gather}\label{eq:torsion_scalars}
    T_{\text{axi}}=a_\mu a^\mu\,,\quad T_{\text{vec}}=v_\mu v^\mu\,,\quad T_{\text{ten}}=t_{\sigma\mu\nu}t^{\sigma\mu\nu}\,,
\end{gather}
which can be used to construct Lagrangians. The general linear combination is known as New General Relativity~\cite{Hayashi:1979qx}, with the Lagrangian density
\begin{equation}\label{eq:lagNGR}
    L_{\text{NGR}}=c_{\text{axi}}T_{\text{axi}}+c_{\text{vec}}T_{\text{vec}}+c_{\text{ten}}T_{\text{ten}}\,.
\end{equation}
Of special interest is the subclass where $c_{\text{ten}}=-c_{\text{vec}}$, which prevents the appearance of ghosts~\cite{VanNieuwenhuizen:1973fi}, and the case where $c_{\text{axi}}=3/2$, $c_{\text{vec}}=-2/3$, $c_{\text{ten}}=2/3$, which corresponds to the teleparallel equivalent (up to a boundary term) of GR (TEGR). We see from this that the former subclass admits a 1-parameter case in which $c_{\text{axi}}=3/2+\delta$, $c_{\text{vec}}=-c_{\text{ten}}=-2/3$, known as 1-parameter NGR (1PNGR), which reduces to TEGR when $\delta=0$~\cite{Hayashi:1979qx}. We also note that the general NGR theory is unstable, but some subclasses, including 1PNGR, have been proven to be stable~\cite{Bahamonde:2024zkb}. As such, we continue with using 1PNGR in this treatment. By varying the corresponding Lagrangian w.r.t. the tetrad, we get the following field equations~\cite{Asukula:2023akj}:
\begin{equation}\label{eq:feq1PNGR}
    E_{\mu\nu}=G_{\mu\nu} + \delta \left( \frac{1}{2} a^\rho a_{(\rho}g_{\mu\nu)} - \frac{4}{9} \epsilon_{\nu\alpha\beta\gamma} a^\alpha {t_\mu}^{\beta\gamma} - \frac{2}{9} \epsilon_{\mu\nu\rho\sigma} a^\rho v^\sigma - \frac{1}{3} \epsilon_{\mu\nu\rho\sigma} \nabla^\rho a^\sigma \right)=\kappa \Theta_{\mu\nu}\,,
\end{equation}
where $G_{\mu\nu}$ is the usual Einstein tensor and $\kappa=8\pi G$ is the Einstein gravitational constant. In the next two sections, we will introduce the teleparallel geometries and energy-momentum tensors that will be used to calculate solutions in section~\ref{sec:cosmofield}.
\section{Bianchi spacetime geometry}\label{sec:geometry}
We start our derivation with a brief overview of the different teleparallel spacetime geometries we will be studying in this article. The most general tetrads describing the three Bianchi cosmologies of type II, III and IX, which admit four-dimensional groups of motions, have been found in~\cite{Hohmann:2023sto} and parametrized in terms of six free functions of the time coordinate. For each type, we briefly review this solution, and then show that it admits a two-dimensional group of time-dependent diffeomorphisms, which allows us to eliminate two of the parameter functions. Further, we introduce a new set of parameter functions, which will be more convenient for our calculations. This is done for type II in section~\ref{ssec:geometryII}, for type III in section~\ref{ssec:geometryIII} and for type IX in section~\ref{ssec:geometryIX}.

\subsection{Type II}\label{ssec:geometryII}
The Bianchi type II tetrad is the most general tetrad such that its metric and affine connection are invariant under the action of the symmetry generators
\begin{equation}\label{eq:vecfII}
X_1 = \partial_y\,, \quad
X_2 = \partial_z\,, \quad
X_3 = z\partial_y - \partial_x\,, \quad
X_4 = z\partial_x + \frac{x^2 - z^2}{2}\partial_y - x\partial_z\,,
\end{equation}
which satisfy the algebra relations
\begin{equation}\label{eq:algII}
[X_1, X_2] = [X_1, X_3] = [X_1, X_4] = 0\,, \quad
[X_2, X_3] = X_1\,, \quad
[X_2, X_4] = -X_3\,, \quad
[X_3, X_4] = X_2\,.
\end{equation}
It has the general form
\begin{subequations}\label{eq:tetII}
\begin{align}
\theta^0 &= [\mathcal{C}_2(1 + x^2 + z^2) + \mathcal{C}_1]\dd t + (\mathcal{C}_3x + \mathcal{C}_4z)\dd x + [\mathcal{C}_6(1 + x^2 + z^2) + \mathcal{C}_5]\dd y + \{[\mathcal{C}_6(1 + x^2 + z^2) + \mathcal{C}_5 - \mathcal{C}_4]x + \mathcal{C}_3z\}\dd z\,,\\
\theta^1 &= 2\mathcal{C}_2x\dd t + \mathcal{C}_3\dd x + 2\mathcal{C}_6x\dd y + (2\mathcal{C}_6x^2 - \mathcal{C}_4)\dd z\,,\\
\theta^2 &= 2\mathcal{C}_2z\dd t + \mathcal{C}_4\dd x + 2\mathcal{C}_6z\dd y + (2\mathcal{C}_6xz + \mathcal{C}_3)\dd z\,,\\
\theta^3 &= [\mathcal{C}_2(1 - x^2 - z^2) - \mathcal{C}_1]\dd t - (\mathcal{C}_3x + \mathcal{C}_4z)\dd x + [\mathcal{C}_6(1 - x^2 - z^2) - \mathcal{C}_5]\dd y + \{[\mathcal{C}_6(1 - x^2 - z^2) - \mathcal{C}_5 + \mathcal{C}_4]x - \mathcal{C}_3z\}\dd z\,,
\end{align}
\end{subequations}
where $\mathcal{C}_i$ are functions of time. The tetrad determinant in this case is
\begin{equation}\label{eq:dettetII}
\det{{\theta^a}_\mu}=2(\mathcal{C}_3^2+\mathcal{C}_4^2)(\mathcal{C}_2\mathcal{C}_5-\mathcal{C}_1\mathcal{C}_6).
\end{equation}
The non-vanishing metric components are
\begin{subequations}\label{eq:metII}
\begin{gather}
g_{tt} = -4\mathcal{C}_1\mathcal{C}_2\,, \quad
g_{ty} = -2(\mathcal{C}_2\mathcal{C}_5 + \mathcal{C}_1\mathcal{C}_6)\,, \quad
g_{xx} = \mathcal{C}_3^2 + \mathcal{C}_4^2\,, \quad
g_{yy} = -4\mathcal{C}_5\mathcal{C}_6\,,\label{eq:metindII}\\
g_{tz} = xg_{ty}\,, \quad
g_{yz} = xg_{yy}\,, \quad
g_{zz} = g_{xx} + x^2g_{yy}\,.
\end{gather}
\end{subequations}
The metric can be greatly simplified by the following reparametrization of the tetrad functions:
\begin{equation}\label{eq:reparII}
\mathcal{C}_1 = \frac{\eta}{2}e^{-\alpha}\,, \quad
\mathcal{C}_2 = \frac{\eta}{2}e^{\alpha}\,, \quad
\mathcal{C}_3 = \mu\cos\gamma\,, \quad
\mathcal{C}_4 = \mu\sin\gamma\,, \quad
\mathcal{C}_5 = \frac{\lambda}{2}e^{-\beta}\,, \quad
\mathcal{C}_6 = -\frac{\lambda}{2}e^{\beta}\,,
\end{equation}
where $(\eta,\alpha,\beta,\gamma,\mu,\lambda)$ are again functions of time. As mentioned at the beginning of this section, we can also eliminate two of these functions by an appropriate coordinate transformation, which leaves the symmetry generators~\eqref{eq:vecfII} invariant and reads
\begin{equation}
\tilde{t} = F(t)\,, \quad
\tilde{x} = x\,, \quad
\tilde{y} = y + G(t)\,, \quad
\tilde{z} = z\,.
\end{equation}
In the new coordinates, the tetrad~\eqref{eq:tetII} retains its form, with the transformed parameter functions
\begin{equation}
\mathcal{C}_1 = \tilde{\mathcal{C}}_1F' + \tilde{\mathcal{C}}_5G'\,, \quad
\mathcal{C}_2 = \tilde{\mathcal{C}}_2F' + \tilde{\mathcal{C}}_6G'\,, \quad
\mathcal{C}_3 = \tilde{\mathcal{C}}_3\,, \quad
\mathcal{C}_4 = \tilde{\mathcal{C}}_4\,, \quad
\mathcal{C}_5 = \tilde{\mathcal{C}}_5\,, \quad
\mathcal{C}_6 = \tilde{\mathcal{C}}_6\,,
\end{equation}
where \(\tilde{\mathcal{C}}_i\) are now functions of \(\tilde{t}\). Hence, we find that the new parameter functions transform as
\begin{gather}
\eta = \sqrt{\tilde{\eta}^2F'^2 - \tilde{\lambda}^2G'^2 + 2\tilde{\eta}\tilde{\lambda}F'G'\sinh(\tilde{\alpha} - \tilde{\beta})}\,, \quad
\alpha = \frac{1}{2}\ln\frac{\tilde{\eta}F'e^{\tilde{\alpha}} - \tilde{\lambda}G'e^{\tilde{\beta}}}{\tilde{\eta}F'e^{-\tilde{\alpha}} + \tilde{\lambda}G'e^{-\tilde{\beta}}}\,, \nonumber\\
\lambda = \tilde{\lambda}\,, \quad
\beta = \tilde{\beta}\,, \quad
\mu = \tilde{\mu}\,, \quad
\gamma = \tilde{\gamma}\,.
\end{gather}
Choosing
\begin{equation}
F' = \frac{1}{\tilde{\eta}\cosh(\tilde{\alpha} - \tilde{\beta})}\,, \quad
G' = \frac{\tanh(\tilde{\alpha} - \tilde{\beta})}{\tilde{\lambda}}\,,
\end{equation}
we can always achieve \(\eta = 1\) and \(\alpha = \beta\). In the following, we will therefore impose this condition on the parameter functions, which eliminates the coordinate freedom. In these coordinates, the independent metric components~\eqref{eq:metindII} become
\begin{gather}\label{eq:metreparII}
g_{tt} = -1\,, \quad
g_{ty} = 0\,, \quad
g_{xx} = \mu^2\,, \quad
g_{yy} = \lambda^2\,,
\end{gather}
while the tetrad determinant simplifies to
\begin{equation}\label{eq:dettet_reparII}
\det{{\theta^a}_\mu}=\lambda\mu^2.
\end{equation}
We note that $\mu$ and $\lambda$ play the role of scale factors. To see this, note that the vector fields \(\partial_t\) and \(X_1 = \partial_y\) are invariant under the action of the symmetry generators~\eqref{eq:vecfII}. While the former is a unit timelike hypersurface normal, \(g(\partial_t, \partial_t) = -1\), the latter is a spacelike vector field which spans the single isotropy direction of this spacetime and satisfies \(g(\partial_y, \partial_y) = \lambda^2\). The induced metric on the orthogonal complement to these vector fields is proportional to \(\mu^2\). Hence, we can interpret \(\lambda\) as a scale factor for the space parallel to \(X_1\) and \(\mu\) for its orthogonal complement.

\subsection{Type III}\label{ssec:geometryIII}
The Bianchi type III symmetry generators read
\begin{equation}\label{eq:vecfIII}
X_1 = \partial_y\,, \quad
X_2 = \partial_z\,, \quad
X_3 = \partial_x - y\partial_y\,, \quad
X_4 = y\partial_x + \frac{1}{2}\left(\frac{e^{-2x}}{1 - n^2} - y^2\right)\partial_y - \frac{ne^{-x}}{1 - n^2}\partial_z\,.
\end{equation}
They satisfy the commutation relations
\begin{equation}\label{eq:algIII}
[X_1, X_2] = [X_2, X_3] = [X_2, X_4] = 0\,, \quad
[X_1, X_3] = -X_1\,, \quad
[X_1, X_4] = X_3\,, \quad
[X_3, X_4] = -X_4\,.
\end{equation}
In this case, the most general tetrad has the form
\begin{subequations}\label{eq:tetIII}
\begin{align}
\theta^0 &= \frac{\mathcal{C}_1}{2\sqrt{2}}\left[\frac{e^{-x}}{\sqrt{1 - n^2}} + \sqrt{1 - n^2}(y^2 + 2)e^x\right]\dd t + \left\{\left[\frac{e^{-x}}{\sqrt{1 - n^2}} - (y^2 + 2)e^x\right]\frac{\mathcal{C}_2}{2\sqrt{2}}- \frac{\mathcal{C}_3y}{\sqrt{2}\sqrt{1 - n^2}}\right\}\dd x\nonumber\\
&\phantom{=}+ \frac{1}{2\sqrt{2}}\left[\sqrt{1 - n^2}(\mathcal{C}_3 + \mathcal{C}_4)(y^2 + 2)e^{2x} - \frac{\mathcal{C}_3 - \mathcal{C}_4}{\sqrt{1 - n^2}} - 2ye^x\mathcal{C}_2\right]\dd y + \frac{\mathcal{C}_4}{2\sqrt{2}n}\left[\frac{e^x}{\sqrt{1 - n^2}} + \sqrt{1 - n^2}(y^2 + 2)e^x\right]\dd z\,,\\
\theta^1 &= \mathcal{C}_5\dd t + \mathcal{C}_6e^x\dd y + \frac{\mathcal{C}_6}{n}\dd z\,,\\
\theta^2 &= -\sqrt{1 - n^2}\mathcal{C}_1ye^x\dd t + \left(\mathcal{C}_2ye^x + \frac{\mathcal{C}_3}{\sqrt{1 - n^2}}\right)\dd x + \left[\mathcal{C}_2e^x - \sqrt{1 - n^2}(\mathcal{C}_3 + \mathcal{C}_4)ye^{2x}\right]\dd y - \frac{\sqrt{1 - n^2}}{n}\mathcal{C}_4ye^x\dd z\,,\\
\theta^3 &= \frac{\mathcal{C}_1}{2\sqrt{2}}\left[\frac{e^{-x}}{\sqrt{1 - n^2}} + \sqrt{1 - n^2}(y^2 - 2)e^x\right]\dd t + \left\{\left[\frac{e^{-x}}{\sqrt{1 - n^2}} - (y^2 - 2)e^x\right]\frac{\mathcal{C}_2}{2\sqrt{2}}- \frac{\mathcal{C}_3y}{\sqrt{2}\sqrt{1 - n^2}}\right\}\dd x\nonumber\\
&\phantom{=}+ \frac{1}{2\sqrt{2}}\left[\sqrt{1 - n^2}(\mathcal{C}_3 + \mathcal{C}_4)(y^2 - 2)e^{2x} - \frac{\mathcal{C}_3 - \mathcal{C}_4}{\sqrt{1 - n^2}} - 2ye^x\mathcal{C}_2\right]\dd y + \frac{\mathcal{C}_4}{2\sqrt{2}n}\left[\frac{e^x}{\sqrt{1 - n^2}} + \sqrt{1 - n^2}(y^2 - 2)e^x\right]\dd z\,.
\end{align}
\end{subequations}
where $\mathcal{C}_i$ are again functions of time. Note that we now have an additional parameter $0<n<1$. Before moving on, we show how to eliminate this parameter. Since $n$ is a constant, we can perform the coordinate transformations $x\rightarrow x-\frac{\ln(1-n^2)}{2}$ and $z\rightarrow \frac{nz}{\sqrt{1-n^2}}$ and also rescale four of the parameter functions as $\mathcal{C}_{2,3,4,6}\rightarrow\mathcal{C}_{2,3,4,6}\sqrt{1-n^2}$. This eliminates $n$ from the tetrad while preserving the base algebra. Now, we continue with this new tetrad. The new determinant is
\begin{equation}\label{eq:dettetIII}
\det{{\theta^a}_\mu}=e^x(\mathcal{C}_2^2+\mathcal{C}_3^2)(\mathcal{C}_4\mathcal{C}_5-\mathcal{C}_1\mathcal{C}_6).
\end{equation}
The non-vanishing metric components are
\begin{subequations}\label{eq:metIII}
\begin{gather}
g_{tt} = \mathcal{C}_5^2 - \mathcal{C}_1^2\,, \quad
g_{tz} = \mathcal{C}_5\mathcal{C}_6 - \mathcal{C}_1\mathcal{C}_4\,, \quad
g_{xx} = \mathcal{C}_2^2 + \mathcal{C}_3^2\,, \quad
g_{zz} = \mathcal{C}_6^2 - \mathcal{C}_4^2\,,\label{eq:metindIII}\\
g_{ty} = g_{tz}e^x\,, \quad
g_{yz} = g_{zz}e^x\,, \quad
g_{yy} = \left(g_{xx} + g_{zz}\right)e^{2x}\,.
\end{gather}
\end{subequations}
Again, we find a convenient reparametrization, which now reads
\begin{equation}\label{eq:reparIII}
\mathcal{C}_1 = \eta\cosh\alpha\,, \quad
\mathcal{C}_2 = \mu\cos\gamma\,, \quad
\mathcal{C}_3 = \mu\sin\gamma\,, \quad
\mathcal{C}_4 = \lambda\sinh\beta\,, \quad
\mathcal{C}_5 = \eta\sinh\alpha\,, \quad
\mathcal{C}_6 = \lambda\cosh\beta\,.
\end{equation}
As for the previously discussed case, the symmetry generators~\eqref{eq:vecfIII} allow for a residual coordinate freedom, which is given by
\begin{equation}
\tilde{t} = F(t)\,, \quad
\tilde{x} = x\,, \quad
\tilde{y} = y\,, \quad
\tilde{z} = z + G(t)\,.
\end{equation}
Again, we find that the tetrad~\eqref{eq:tetIII} retains its form, while the parameter functions now transform as
\begin{equation}
\mathcal{C}_1 = \tilde{\mathcal{C}}_1F' + \tilde{\mathcal{C}}_4G'\,, \quad
\mathcal{C}_2 = \tilde{\mathcal{C}}_2\,, \quad
\mathcal{C}_3 = \tilde{\mathcal{C}}_3\,, \quad
\mathcal{C}_4 = \tilde{\mathcal{C}}_4\,, \quad
\mathcal{C}_5 = \tilde{\mathcal{C}}_5F' + \tilde{\mathcal{C}}_6G'\,, \quad
\mathcal{C}_6 = \tilde{\mathcal{C}}_6\,.
\end{equation}
For the reparametrization, we find the transformation
\begin{gather}
\eta = \sqrt{\tilde{\eta}^2F'^2 - \tilde{\lambda}^2G'^2 - 2\tilde{\eta}\tilde{\lambda}F'G'\sinh(\tilde{\alpha} - \tilde{\beta})}\,, \quad
\alpha = \operatorname{arctanh}\frac{\tilde{\eta}F'\sinh\tilde{\alpha} + \tilde{\lambda}G'\cosh\tilde{\beta}}{\tilde{\eta}F'\cosh\tilde{\alpha} + \tilde{\lambda}G'\sinh\tilde{\beta}}\,, \nonumber\\
\lambda = \tilde{\lambda}\,, \quad
\beta = \tilde{\beta}\,, \quad
\mu = \tilde{\mu}\,, \quad
\gamma = \tilde{\gamma}\,.
\end{gather}
Similarly to the previous case, we can now choose
\begin{equation}
F' = \frac{1}{\tilde{\eta}\cosh(\tilde{\alpha} - \tilde{\beta})}\,, \quad
G' = -\frac{\tanh(\tilde{\alpha} - \tilde{\beta})}{\tilde{\lambda}}\,,
\end{equation}
in order to achieve \(\eta = 1\) and \(\alpha = \beta\). This yields the independent metric components~\eqref{eq:metindIII}
\begin{gather}\label{eq:metreparIII}
g_{tt} = -1\,, \quad
g_{tz} = 0\,, \quad
g_{xx} = \mu^2\,, \quad
g_{zz} = \lambda^2\,,
\end{gather}
whereas the tetrad determinant now becomes
\begin{equation}\label{eq:dettet_reparIII}
\det{{\theta^a}_\mu}=e^x\lambda\mu^2\,.
\end{equation}
Again, $\mu$ and $\lambda$ act as scale factors. This can be seen by noting that \(\partial_t\) and \(X_2 = \partial_z\) are invariant under the action of the generating vector fields~\eqref{eq:vecfIII}, with \(g(\partial_t, \partial_t) = -1\) and \(g(\partial_z, \partial_z) = \lambda^2\). Hence, we find again that \(\lambda\) measures the expansion parallel to the isotropy direction, while \(\mu\) acts as a scale factor on the transverse directions.

\subsection{Type IX}\label{ssec:geometryIX}
We finally come to the Bianchi type IX. The symmetry-generating vector fields are given by
\begin{equation}\label{eq:vecfIX}
X_1 = \partial_y\,, \quad
X_2 = n\frac{\sin y}{\sin x}\partial_z + \cos y\partial_x - \frac{\sin y}{\tan x}\partial_y\,, \quad
X_3 = n\frac{\cos y}{\sin x}\partial_z - \sin y\partial_x - \frac{\cos y}{\tan x}\partial_y\,, \quad
X_4 = \partial_z\,,
\end{equation}
which leads to the commutation relations
\begin{equation}\label{eq:algIX}
[X_1, X_2] = X_3\,, \quad
[X_2, X_3] = X_1\,, \quad
[X_3, X_1] = X_2\,, \quad
[X_1, X_4] = [X_2, X_4] = [X_3, X_4] = 0\,.
\end{equation}
The most general tetrad has the form
\begin{subequations}\label{eq:tetIX}
\begin{align}
\theta^0 &= \mathcal{C}_1\dd t + n\mathcal{C}_2\cos x\dd y + \mathcal{C}_2\dd z\,,\\
\theta^1 &= \mathcal{C}_3\cos x\dd t - \mathcal{C}_4\sin x\dd x + (n\mathcal{C}_5\cos^2x - \mathcal{C}_6\sin^2x)\dd y + \mathcal{C}_5\cos x\dd z\,,\\
\theta^2 &= \mathcal{C}_3\sin x\sin y\dd t + (\mathcal{C}_4\cos x\sin y - \mathcal{C}_6\cos y)\dd x + \mathcal{C}_5\sin x\sin y\dd z + [(n\mathcal{C}_5 + \mathcal{C}_6)\cos x\sin y + \mathcal{C}_4\cos y]\sin x\dd y\,,\\
\theta^3 &= \mathcal{C}_3\sin x\cos y\dd t + (\mathcal{C}_4\cos x\cos y + \mathcal{C}_6\sin y)\dd x + \mathcal{C}_5\sin x\cos y\dd z + [(n\mathcal{C}_5 + \mathcal{C}_6)\cos x\cos y - \mathcal{C}_4\sin y]\sin x\dd y\,,
\end{align}
\end{subequations}
where $\mathcal{C}_i$ are again functions of time. Note that we now have an additional parameter $n\neq0$. Similarly to the previous type, we can eliminate this parameter. Since $n$ is a constant, we can perform a simple coordinate transformation $z\rightarrow nz$ and also rescale two of the parameter functions as $\mathcal{C}_2\rightarrow\mathcal{C}_2/n$ and $\mathcal{C}_5\rightarrow\mathcal{C}_5/n$. This eliminates $n$ from the tetrad entirely, with the result being identical to what one would get by setting $n=1$ in eq. \eqref{eq:tetIX}. With this, the tetrad determinant is
\begin{equation}\label{eq:dettetIX}
\det{{\theta^a}_\mu}=(\mathcal{C}_4^2+\mathcal{C}_6^2)(\mathcal{C}_2\mathcal{C}_3-\mathcal{C}_1\mathcal{C}_5)\sin x.
\end{equation}
Moving on, the non-vanishing metric components are
\begin{subequations}\label{eq:metIX}
\begin{gather}
g_{tt} = \mathcal{C}_3^2 - \mathcal{C}_1^2\,, \quad
g_{tz} = \mathcal{C}_3\mathcal{C}_5 - \mathcal{C}_1\mathcal{C}_2\,, \quad
g_{xx} = \mathcal{C}_4^2 + \mathcal{C}_6^2\,, \quad
g_{zz} = \mathcal{C}_5^2 - \mathcal{C}_2^2\,,\label{eq:metindIX}\\
g_{ty} = g_{tz}\cos x\,, \quad
g_{yz} = g_{zz}\cos x\,, \quad
g_{yy} = g_{xx}\sin^2x + g_{zz}\cos^2x\,.
\end{gather}
\end{subequations}
Again, we perform a reparametrization, which is now given by
\begin{equation}\label{eq:reparIX}
\mathcal{C}_1 = \eta\cosh\alpha\,, \quad
\mathcal{C}_2 = \lambda\sinh\beta\,, \quad
\mathcal{C}_3 = \eta\sinh\alpha\,, \quad
\mathcal{C}_4 = \mu\cos\gamma\,, \quad
\mathcal{C}_5 = \lambda\cosh\beta\,, \quad
\mathcal{C}_6 = \mu\sin\gamma\,.
\end{equation}
Finally, we once again perform the coordinate transformation
\begin{equation}
\tilde{t} = F(t)\,, \quad
\tilde{x} = x\,, \quad
\tilde{y} = y\,, \quad
\tilde{z} = z + G(t)\,,
\end{equation}
from which we read off the transformations
\begin{equation}
\mathcal{C}_1 = \tilde{\mathcal{C}}_1F' + \tilde{\mathcal{C}}_2G'\,, \quad
\mathcal{C}_2 = \tilde{\mathcal{C}}_2\,, \quad
\mathcal{C}_3 = \tilde{\mathcal{C}}_3F' + \tilde{\mathcal{C}}_5G'\,, \quad
\mathcal{C}_4 = \tilde{\mathcal{C}}_4\,, \quad
\mathcal{C}_5 = \tilde{\mathcal{C}}_5\,, \quad
\mathcal{C}_6 = \tilde{\mathcal{C}}_6\
\end{equation}
of the parameter functions, from which we find that the new parameter functions transform as
\begin{gather}
\eta = \sqrt{\tilde{\eta}^2F'^2 - \tilde{\lambda}^2G'^2 - 2\tilde{\eta}\tilde{\lambda}F'G'\sinh(\tilde{\alpha} - \tilde{\beta})}\,, \quad
\alpha = \artanh\frac{\tilde{\eta}F'\sinh\tilde{\alpha} + \tilde{\lambda}G'\cosh\tilde{\beta}}{\tilde{\eta}F'\cosh\tilde{\alpha} + \tilde{\lambda}G'\sinh\tilde{\beta}}\,, \nonumber\\
\lambda = \tilde{\lambda}\,, \quad
\beta = \tilde{\beta}\,, \quad
\mu = \tilde{\mu}\,, \quad
\gamma = \tilde{\gamma}\,.
\end{gather}
By choosing
\begin{equation}
F' = \frac{1}{\tilde{\eta}\cosh(\tilde{\alpha} - \tilde{\beta})}\,, \quad
G' = -\frac{\tanh(\tilde{\alpha} - \tilde{\beta})}{\tilde{\lambda}}\,,
\end{equation}
we achieve \(\eta = 1\) and \(\alpha = \beta\), as for the previous cases. In this case, we find the independent metric components~\eqref{eq:metindIX}
\begin{equation}\label{eq:metreparIX}
g_{tt} = -1\,, \quad
g_{tz} = 0\,, \quad
g_{xx} = \mu^2\,, \quad
g_{zz} = \lambda^2\,,
\end{equation}
as well as the new tetrad determinant
\begin{equation}\label{eq:dettet_reparIX}
\det{{\theta^a}_\mu}=\lambda\mu^2\sin x\,.
\end{equation}
Again, $\mu$ and $\lambda$ act as scale factors, where now \(\partial_t\) and \(X_4 = \partial_z\) are invariant under the action of the generating vector fields~\eqref{eq:vecfIX}, with \(g(\partial_t, \partial_t) = -1\) and \(g(\partial_z, \partial_z) = \lambda^2\). We thus conclude that \(\lambda\) again measures the expansion parallel to the isotropy direction, while \(\mu\) acts as a scale factor in the transverse directions.

\section{Energy-momentum tensor of an anisotropic fluid}\label{sec:enmom}
We now discuss the most general energy-momentum sources which are compatible with the four-dimensional spacetime symmetry groups outlined in the previous section. First, we use the invariance condition
\begin{equation}\label{eq:enmominv}
(\mathcal{L}_{X_i}\Theta)_{\mu\nu} = 0
\end{equation}
in order to determine the most general, invariant, symmetric tensor \(\Theta_{\mu\nu}\). In analogy to the metric, one finds that it depends on four free functions of time. We then make use of the parameter functions introduced for constructing the tetrad in order to introduce a new parametrization for these matter variables. The key idea here is to raise one index of the energy-momentum tensor with the metric to obtain an endomorphism \(\Theta^{\mu}{}_{\nu}\) of the tangent bundle and determine its eigenvectors and eigenvalues. From the fact that \(\Theta_{\mu\nu}\) is symmetric follows that the eigenvectors which belong to distinct eigenvalues are orthogonal. Further, it follows from the invariance under the action of the symmetry group, which entails a one-dimensional isotropy group at each point, that two eigenvalues are identical and belong to spacelike eigenvectors. The remaining two eigenvalues belong to one timelike eigenvector field \(U^{\mu}\) and one spacelike eigenvector field \(W^{\mu}\), both of which are invariant under the action of the symmetry group. Choosing these to be normalized,
\begin{equation}
U^{\mu}U_{\mu} = -1\,, \quad
W^{\mu}W_{\mu} = 1\,, \quad
U^{\mu}W_{\mu} = 0\,,
\end{equation}
and taking \(U^{\mu}\) future-pointing, we can thus write the energy-momentum tensor as an anisotropic fluid of the form
\begin{equation}\label{eq:enmom_tensor_general}
\Theta_{\mu\nu} = g_{\mu\nu}p_{\perp} + (\rho + p_{\perp})U_{\mu}U_{\nu} + (p_{\parallel} - p_{\perp})W_{\mu}W_{\nu}\,,
\end{equation}
where we introduced the energy density \(\rho\) and the collinear and transverse pressure \(p_{\parallel}\) and \(p_{\perp}\). One easily checks that this expression satisfies the eigenvector relations
\begin{equation}
\Theta^{\mu}{}_{\nu}U^{\nu} = -\rho U^{\nu}\,, \quad
\Theta^{\mu}{}_{\nu}W^{\nu} = p_{\parallel}W^{\nu}\,, \quad
\Theta^{\mu}{}_{\nu}V^{\nu} = p_{\perp}V^{\nu}\,,
\end{equation}
where \(V^{\mu}\) is any vector field in the orthogonal complement of \(U^{\mu}\) and \(W^{\nu}\). If we fix an orthonormal basis, comprised of a future-pointing, timelike vector field \(T^{\mu}\) and a spacelike vector field \(S^{\mu}\), of the invariant subbundle of the tangent bundle spanned by \(U^{\mu}\) and \(W^{\mu}\), we can uniquely express the latter in the form
\begin{equation}
U^{\mu} = T^{\mu}\cosh\nu + S^{\mu}\sinh\nu\,, \quad
W^{\mu} = T^{\mu}\sinh\nu + S^{\mu}\cosh\nu\,,
\end{equation}
and so we see that they are fully determined by the invariance under the symmetry group and the orthonormality conditions, up to a single function \(\nu\) of time, which can be interpreted as the rapidity of the fluid in the reference frame established by \(T^{\mu}\) and \(S^{\mu}\). The four functions \(\rho, p_{\parallel}, p_{\perp}, \nu\) of time are then used to parametrize the most general energy-momentum tensor compatible with the symmetry group. We will now conduct this procedure for type II in section~\ref{ssec:enmomII}, for type III in section~\ref{ssec:enmomIII} and for type IX in section~\ref{ssec:enmomIX}. Finally, we also discuss the covariant energy-momentum conservation equations.

\subsection{Type II}\label{ssec:enmomII}
By solving the invariance condition~\eqref{eq:enmominv} for the vector fields~\eqref{eq:vecfII}, one finds that the most general energy-momentum tensor, which is invariant under the action of the symmetry generating vector fields, has the non-vanishing components
\begin{subequations}\label{eq:enmomII}
\begin{gather}
\Theta_{tt} = \mathcal{M}_1\,, \quad
\Theta_{ty} = \mathcal{M}_2\,, \quad
\Theta_{xx} = \mathcal{M}_3\,, \quad
\Theta_{yy} = \mathcal{M}_4\,,\\
\Theta_{tz} = x\Theta_{ty}\,, \quad
\Theta_{yz} = x\Theta_{yy}\,, \quad
\Theta_{zz} = \Theta_{xx} + x^2\Theta_{yy}\,,
\end{gather}
\end{subequations}
where \(\mathcal{M}_i\) are four functions of time. To write the energy-momentum tensor in the form~\eqref{eq:enmom_tensor_general}, we choose the orthonormal vectors
\begin{subequations}
\begin{align}
T^{\mu}\partial_{\mu} &= \frac{(\mathcal{C}_5 - \mathcal{C}_6)\partial_t - (\mathcal{C}_1 - \mathcal{C}_2)\partial_y}{2(\mathcal{C}_2\mathcal{C}_5 - \mathcal{C}_1\mathcal{C}_6)} = \frac{\lambda\cosh\beta\partial_t + \eta\sinh\alpha\partial_y}{\eta\lambda\cosh(\alpha - \beta)}\,,\\
S^{\mu}\partial_{\mu} &= \frac{(\mathcal{C}_5 + \mathcal{C}_6)\partial_t - (\mathcal{C}_1 + \mathcal{C}_2)\partial_y}{2(\mathcal{C}_2\mathcal{C}_5 - \mathcal{C}_1\mathcal{C}_6)} = -\frac{\lambda\sinh\beta\partial_t + \eta\cosh\alpha\partial_y}{\eta\lambda\cosh(\alpha - \beta)}\,.
\end{align}
\end{subequations}
Lowering the index with the metric yields the normalized covectors
\begin{subequations}
\begin{align}
T_{\mu}\dd x^{\mu} &= -(\mathcal{C}_1 + \mathcal{C}_2)\dd t - (\mathcal{C}_5 + \mathcal{C}_6)(\dd y + x\,\dd z) = -\eta\cosh\alpha\,\dd t + \lambda\sinh\beta(\dd y + x\,\dd z)\,,\\
S_{\mu}\dd x^{\mu} &= -(\mathcal{C}_1 - \mathcal{C}_2)\dd t - (\mathcal{C}_5 - \mathcal{C}_6)(\dd y + x\,\dd z) = \eta\sinh\alpha\,\dd t - \lambda\cosh\beta(\dd y + x\,\dd z)\,.
\end{align}
\end{subequations}
The normalized eigenvectors of the energy-momentum tensor then read
\begin{subequations}\label{eq:enmom_vecII}
\begin{align}
U_{\mu}\dd x^{\mu} &= -\eta\cosh(\nu - \alpha)\,\dd t - \lambda\sinh(\nu - \beta)(\dd y + x\,\dd z)\,,\\
W_{\mu}\dd x^{\mu} &= -\eta\sinh(\nu - \alpha)\,\dd t - \lambda\cosh(\nu - \beta)(\dd y + x\,\dd z)\,.
\end{align}
\end{subequations}
Note that we have kept the coordinates arbitrary so far. In the following, we will fix the coordinates such that \(\eta = 1\) and \(\alpha = \beta\), and introduce the abbreviation \(\tilde{\nu} = \nu - \beta\) to simplify the calculation. We can then read off the components
\begin{equation}
\mathcal{M}_1 = \rho\cosh^2\tilde{\nu}\ + p_{\parallel}\sinh^2\tilde{\nu}\,, \quad
\mathcal{M}_2 = -\lambda\frac{\rho + p_{\parallel}}{2}\sinh(2\tilde{\nu})\,, \quad
\mathcal{M}_3 = \mu^2p_{\perp}\,, \quad
\mathcal{M}_4 = \lambda^2(\rho\sinh^2\tilde{\nu}\ + p_{\parallel}\cosh^2\tilde{\nu})\,.
\end{equation}
It is also useful to calculate the energy-momentum conservation equations. By taking the energy-momentum tensor with components~\eqref{eq:enmomII} and the covariant derivative with the Levi-Civita connection of the metric~\eqref{eq:metindII} and then calculating $\nabla_\mu T^{\mu\nu}$ directly, we get:
\begin{equation}
\mathcal{M}_1'+\frac{(\mathcal{M}_4+\mathcal{M}_1\lambda^2)\lambda'}{\lambda^3}+\frac{2(\mathcal{M}_3+\mathcal{M}_1\mu^2)\mu'}{\mu^3}=0\,,\quad \frac{\mathcal{M}_2\lambda'}{\lambda^3}=0\,.
\end{equation}
The second equation can be solved by setting $\mathcal{M}_2=0$, which implies either $\tilde\nu=0$ or $\rho=-p_\parallel$. As we wish to investigate only non-exotic matter-energy, we discard the latter solution in this work. This also simplifies the first equation in terms of $\rho$, $p_\parallel$ and $p_\perp$:
\begin{equation}\label{eq:enmomconII}
    \frac{\rho'}{\rho}=-\left[\left(1+\frac{p_\parallel}{\rho}\right)\frac{\lambda'}{\lambda}+2\left(1+\frac{p_\perp}{\rho}\right)\frac{\mu'}{\mu}\right]\,.
\end{equation}

\subsection{Type III}\label{ssec:enmomIII}
We now insert the vector fields~\eqref{eq:vecfIII} (accounting for the elimination of $n$ as described in the previous section) into the invariance condition~\eqref{eq:enmominv}. In this case, one finds that the most general solution has the non-vanishing components
\begin{subequations}\label{eq:enmomIII}
\begin{gather}
\Theta_{tt} = \mathcal{M}_1\,, \quad
\Theta_{tz} = \mathcal{M}_2\,, \quad
\Theta_{xx} = \mathcal{M}_3\,, \quad
\Theta_{zz} = \mathcal{M}_4\,,\\
\Theta_{ty} = \Theta_{tz}e^x\,, \quad
\Theta_{yz} = \Theta_{zz}e^x\,, \quad
\Theta_{yy} = \left(\Theta_{xx} + \Theta_{zz}\right)e^{2x}\,,
\end{gather}
\end{subequations}
where again \(\mathcal{M}_i\) are four functions of time. Also in this case, we write the energy-momentum tensor as an anisotropic fluid of the form~\eqref{eq:enmom_tensor_general}, where now we choose the orthonormal vectors
\begin{subequations}
\begin{align}
T^{\mu}\partial_{\mu} &= \frac{\mathcal{C}_6\partial_t - \mathcal{C}_5\partial_z}{\mathcal{C}_1\mathcal{C}_6 - \mathcal{C}_4\mathcal{C}_5} = \frac{\lambda\cosh\beta\partial_t - \eta\sinh\alpha\partial_z}{\eta\lambda\cosh(\alpha - \beta)}\,,\\
S^{\mu}\partial_{\mu} &= \frac{\mathcal{C}_1\partial_z - \mathcal{C}_4\partial_t}{\mathcal{C}_1\mathcal{C}_6 - \mathcal{C}_4\mathcal{C}_5} = \frac{-\lambda\sinh\beta\partial_t + \eta\cosh\alpha\partial_z}{\eta\lambda\cosh(\alpha - \beta)}\,,
\end{align}
\end{subequations}
from which we then obtain the normalized covectors
\begin{subequations}
\begin{align}
T_{\mu}\dd x^{\mu} &= -\mathcal{C}_1\dd t - \mathcal{C}_4\left(e^x\dd y + \dd z\right) = -\eta\cosh\alpha\dd t - \lambda\sinh\beta\left(e^x\dd y + \dd z\right)\,,\\
S_{\mu}\dd x^{\mu} &= \mathcal{C}_5\dd t + \mathcal{C}_6\left(e^x\dd y + \dd z\right) = \eta\sinh\alpha\dd t + \lambda\cosh\beta\left(e^x\dd y + \dd z\right)\,.
\end{align}
\end{subequations}
The normalized eigenvectors of the energy-momentum tensor now read
\begin{subequations}\label{eq:enmom_vecIII}
\begin{align}
U_{\mu}\dd x^{\mu} &= -\eta\cosh(\nu - \alpha)\dd t + \lambda\sinh(\nu - \beta)\left(e^x\dd y + \dd z\right)\,,\\
W_{\mu}\dd x^{\mu} &= -\eta\sinh(\nu - \alpha)\dd t + \lambda\cosh(\nu - \beta)\left(e^x\dd y + \dd z\right)\,.
\end{align}
\end{subequations}
Also in this case, we have kept the coordinates arbitrary so far. Fixing the coordinates such that \(\eta = 1\) and \(\alpha = \beta\), and introducing the abbreviation \(\tilde{\nu} = \nu - \beta\), we find the components
\begin{equation}
\mathcal{M}_1 = \rho\cosh^2\tilde{\nu}\ + p_{\parallel}\sinh^2\tilde{\nu}\,, \quad
\mathcal{M}_2 = -\lambda\frac{\rho + p_{\parallel}}{2}\sinh(2\tilde{\nu})\,, \quad
\mathcal{M}_3 = \mu^2p_{\perp}\,, \quad
\mathcal{M}_4 = \lambda^2(\rho\sinh^2\tilde{\nu}\ + p_{\parallel}\cosh^2\tilde{\nu})\,.
\end{equation}
The energy momentum conservation equations, in fact, turn out to be the same as for type II (eq.~\eqref{eq:enmomconII}), so we can take $\tilde\nu=0$ to solve one equation and simplify the other into
\begin{equation}\label{eq:enmomconIII}
    \frac{\rho'}{\rho}=-\left[\left(1+\frac{p_\parallel}{\rho}\right)\frac{\lambda'}{\lambda}+2\left(1+\frac{p_\perp}{\rho}\right)\frac{\mu'}{\mu}\right]\,.
\end{equation}

\subsection{Type IX}\label{ssec:enmomIX}
Finally, we solve the invariance condition~\eqref{eq:enmominv} for the vector fields~\eqref{eq:vecfIX} (with $n=1$). In this case, one finds that the most general energy-momentum tensor satisfying this condition has the non-vanishing components
\begin{subequations}\label{eq:enmomIX}
\begin{gather}
\Theta_{tt} = \mathcal{M}_1\,, \quad
\Theta_{tz} = \mathcal{M}_2\,, \quad
\Theta_{xx} = \mathcal{M}_3\,, \quad
\Theta_{zz} = \mathcal{M}_4\,,\\
\Theta_{ty} = \Theta_{tz}\cos x\,, \quad
\Theta_{yz} = \Theta_{zz}\cos x\,, \quad
\Theta_{yy} = \Theta_{xx}\sin^2x + \Theta_{zz}\cos^2x
\end{gather}
\end{subequations}
in terms of four functions \(\mathcal{M}_i\) of time. Again aiming to write the energy-momentum tensor in the form~\eqref{eq:enmom_tensor_general}, we now choose the orthonormal vectors
\begin{subequations}
\begin{align}
T^{\mu}\partial_{\mu} &= \frac{\mathcal{C}_5\partial_t - \mathcal{C}_3\partial_z}{\mathcal{C}_1\mathcal{C}_5 - \mathcal{C}_2\mathcal{C}_3} = \frac{\lambda\cosh\beta\partial_t - \eta\sinh\alpha\partial_z}{\eta\lambda\cosh(\alpha - \beta)}\,,\\
S^{\mu}\partial_{\mu} &= \frac{\mathcal{C}_1\partial_z - \mathcal{C}_2\partial_t}{\mathcal{C}_1\mathcal{C}_5 - \mathcal{C}_2\mathcal{C}_3} = \frac{-\lambda\sinh\beta\partial_t + \eta\cosh\alpha\partial_z}{\eta\lambda\cosh(\alpha - \beta)}\,,
\end{align}
\end{subequations}
so that by lowering the index with the metric, we obtain the normalized covectors
\begin{subequations}
\begin{align}
T_{\mu}\dd x^{\mu} &= -\mathcal{C}_1\dd t - \mathcal{C}_2(\cos x\,\dd y + \dd z) = -\eta\cosh\alpha\dd t - \lambda\sinh\beta(\cos x\,\dd y + \dd z)\,,\\
S_{\mu}\dd x^{\mu} &= \mathcal{C}_3\dd t + \mathcal{C}_5(\cos x\,\dd y + \dd z) = \eta\sinh\alpha\dd t + \lambda\cosh\beta(\cos x\,\dd y + \dd z)\,.
\end{align}
\end{subequations}
The normalized eigenvectors of the energy-momentum tensor then read
\begin{subequations}\label{eq:enmom_vecIX}
\begin{align}
U_{\mu}\dd x^{\mu} &= -\eta\cosh(\nu - \alpha)\dd t + \lambda\sinh(\nu - \beta)(\cos x\,\dd y + \dd z)\,,\\
W_{\mu}\dd x^{\mu} &= -\eta\sinh(\nu - \alpha)\dd t + \lambda\cosh(\nu - \beta)(\cos x\,\dd y + \dd z)\,.
\end{align}
\end{subequations}
Choosing particular coordinates such that \(\eta = 1\) and \(\alpha = \beta\) and introducing the abbreviation \(\tilde{\nu} = \nu - \beta\) as before, we find the components
\begin{equation}
\mathcal{M}_1 = \rho\cosh^2\tilde{\nu}\ + p_{\parallel}\sinh^2\tilde{\nu}\,, \quad
\mathcal{M}_2 = -\lambda\frac{\rho + p_{\parallel}}{2}\sinh(2\tilde{\nu})\,, \quad
\mathcal{M}_3 = \mu^2p_{\perp}\,, \quad
\mathcal{M}_4 = \lambda^2(\rho\sinh^2\tilde{\nu}\ + p_{\parallel}\cosh^2\tilde{\nu})\,.
\end{equation}
Once again, the energy-momentum conservation equations are unchanged from type II, so we can take $\tilde\nu=0$ and get
\begin{equation}\label{eq:enmomconIX}
    \frac{\rho'}{\rho}=-\left[\left(1+\frac{p_\parallel}{\rho}\right)\frac{\lambda'}{\lambda}+2\left(1+\frac{p_\perp}{\rho}\right)\frac{\mu'}{\mu}\right]\,.
\end{equation}

\section{Cosmological field equations}\label{sec:cosmofield}
With the preliminary calculations at hand, we can now apply the general field equations of NGR displayed in section~\ref{sec:ngr} to the tetrads derived in section~\ref{sec:geometry} and the energy-momentum tensors found in section~\ref{sec:enmom}. Note that these field equations are underdetermined and must be supplemented by an equation of state for the anisotropic fluid in order to obtain a complete system. Here we use a barotropic equation of state given by \(p_{\parallel} = w_{\parallel}\rho\) and \(p_{\perp} = w_{\perp}\rho\), with constants \(w_{\parallel}\) and \(w_{\perp}\), where we pay particular attention to the values \(0\) (dust) and \(1/3\) (radiation). Substituting this equation of state into the energy-momentum conservation equations found in section~\ref{sec:enmom}, we find that for all three Bianchi models the conservation equation becomes
\begin{equation}
\frac{\rho'}{\rho}=-\left[(1+w_\parallel)\frac{\lambda'}{\lambda}+2(1+w_\perp)\frac{\mu'}{\mu}\right]\,,
\end{equation}
from which we see that higher values of \(w_\parallel\) and \(w_\perp\) correspond to a faster decrease of \(\rho\) as the universe expands. Hence, we expect that radiation (\(w = 1/3\)) dominates energy-momentum for small scale factors and dust (\(w = 0\)) dominates for large scale factors. We finally show how the cosmological field equations can be formulated as a dynamical system by introducing auxiliary variables that reduce the second-order time derivatives to first-order derivatives of the new variables, and then solving the system for these time derivatives. This is done for type II in section~\ref{ssec:cosmofieldII}, for type III in section~\ref{ssec:cosmofieldIII} and for type IX in section~\ref{ssec:cosmofieldIX}.

Using the parametrizations and coordinates derived in section \ref{sec:geometry} and the requirement $\tilde\nu=\nu-\beta=0$ derived in section \ref{sec:enmom}, we can significantly simplify the field equations for each type. In what follows, we present the field equations in covariant form, as in~\eqref{eq:feq1PNGR}. We will see that each type has 5 independent equations: 3 symmetric components $E_{(\mu\nu)}$ on the main diagonal that link to the matter variables $\rho$, $p_\parallel$ and $p_\perp$ derived in the previous section, and 2 antisymmetric components $E_{[\mu\nu]}=0$. We first solve the antisymmetric field equations simultaneously for $\gamma'$ and $\beta'$ (or these functions directly); then, we solve the remaining symmetric field equations for $\rho$, $\mu''$, and $\lambda''$. Finally, we present the results as first-order dynamical systems by defining the auxiliary variables \(\psi = \mu'\) and \(\phi = \lambda'\).
\subsection{Type II}\label{ssec:cosmofieldII}
The symmetric field equations for type II are:
\begin{subequations}\label{eq:feqsymII}
\begin{align}
    E_{(00)}=&\frac{\left(2\delta-9\right)\lambda^3-16\delta\lambda^2\mu e^\beta\sin\gamma+72\mu^3\lambda'\mu'+4\lambda\left[2\delta\mu^4(\gamma')^2+9\mu^2(\mu')^2\right]}{36\lambda\mu^4}=\kappa\rho\,,\\
    E_{(11)}=&\frac{\left(2\delta-9\right)\lambda^3-8\delta\lambda^2\mu e^\beta\sin\gamma-36\mu^3\left(\lambda'\mu'+\mu\lambda''\right)+4\lambda\mu^3\left[2\delta\gamma'\left(2e^2\beta\sin\gamma+\mu\gamma'\right)-9\mu''\right]}{36\lambda\mu^2}=\kappa w_\perp\mu^2\rho\,,\\
    E_{(22)}=&\lambda^2\frac{3\left(9-2\delta\right)\lambda^2+32\delta\lambda\mu e^\beta\sin\gamma+4\mu^2\left\{2\mu\left[\delta\gamma'\left(4e^\beta\sin\gamma+\mu\gamma'\right)-9\mu''\right]-9(\mu')^2\right\}}{36\mu^4}=\kappa w_\parallel\lambda^2\rho\,,
\end{align}
\end{subequations}
and the antisymmetric equations are:
\begin{subequations}\label{eq:feqantisymII}
\begin{align}
    E_{[13]}=&\frac{2\delta\lambda e^\beta\sin\gamma\left(\lambda+2\mu^2\gamma'\right)}{9\mu^3}=0\,,\\
    E_{[24]}=&\frac{2\delta\left\{\lambda^2e^\beta\cos\gamma-2\mu\left[\mu\lambda'+\lambda\left(\mu\beta'+\mu'\right)\right]e^\beta\sin\gamma-\mu^2\left[\gamma'\left(\mu\lambda'+2\lambda\mu'\right)+\lambda\mu\gamma''\right]\right\}}{9\lambda\mu}=0\,.
\end{align}
\end{subequations}
First, we can solve $E_{[13]}=0$. We see that there are two branches of solutions, either $\gamma=N\pi,\ N\in\mathbb{Z}$ or:
\begin{equation}\label{eq:gsolII}
    \gamma'=-\frac{\lambda}{2\mu^2}\,.
\end{equation}
Let's look at the first branch. By substituting $\gamma=N\pi$ into $E_{[24]}$, we get
\begin{equation}
    E_{[24]}\propto\frac{\delta\lambda e^\beta}{\mu}\,,
\end{equation}
which is always non-zero (assuming $\delta\neq0$). Therefore, we discard this branch and move on to the second one. By taking the derivative of eq.~\eqref{eq:gsolII} and substituting it into $E_{[24]}$, we get an equation linear in $\beta'$ with the solution:
\begin{equation}\label{eq:bsolII}
    \beta'=\frac{\lambda'\lambda\mu e^{-\beta}\csc\gamma+\lambda^2\cot\gamma-2\mu\left(\lambda'\mu+\mu'\lambda\right)}{2\lambda\mu^2}\,.
\end{equation}
Now we note that the symmetric field equations are linear in $\rho$, $\lambda''$ and $\mu''$. By solving them simultaneously, we get:
\begin{subequations}\label{eq:symsolII}
\begin{align}
    \mu''=&\frac{16 e^\beta\delta\lambda^2\mu\left(w_\parallel+1\right)\sin\gamma+\lambda^3\left[9\left(w_\parallel+3\right)-4 \delta\left(w_\parallel+1\right)\right]-36\lambda\mu^2\left(w_\parallel+1\right)(\mu')^2-72\mu^3w_\parallel\mu'\lambda'}{72\lambda\mu^3}\,,\\
    \lambda''=&\frac{1}{72\mu^4}\left\{-16e^{\beta}\delta\lambda^2\mu\sin\gamma\left(w_\parallel-2 w_\perp+3\right)+\lambda^3\left[4\delta\left(w_\parallel-2w_\perp+3\right)-9\left(w_\parallel-2w_\perp+5\right)\right]\right.\\
    &\left.+36\lambda\mu^2(\mu')^2\left(w_\parallel-2w_\perp+1\right)+72\mu^3\mu'\lambda'\left(w_\parallel-2 w_\perp-1\right)\right\}\,,\nonumber\\
    \kappa\rho=&\frac{\left(4\delta-9\right)\lambda^3-4\mu\left[4\delta\lambda^2 e^\beta\sin\gamma+9\mu\mu'\left(2\mu\lambda'+\lambda\mu'\right)\right]}{36\lambda\mu^4}\,.
\end{align}
\end{subequations}
Finally, we make this into a first-order dynamical system by introducing \(\psi = \mu'\) and \(\phi = \lambda'\). The full system is:
\begin{subequations}\label{eq:dynsysII}
\begin{align}
    \gamma'=&-\frac{\lambda}{2\mu^2}\,,\\
    \beta'=&\frac{\lambda\mu\phi e^{-\beta}\csc\gamma+\lambda^2\cot\gamma-2\mu\left(\mu\phi+\lambda\psi\right)}{2\lambda\mu^2}\,,\\
    \psi'=&\frac{16 e^\beta\delta\lambda^2\mu\left(w_\parallel+1\right)\sin\gamma+\lambda^3\left[9\left(w_\parallel+3\right)-4 \delta\left(w_\parallel+1\right)\right]-36\lambda\mu^2\left(w_\parallel+1\right)\psi^2-72\mu^3w_\parallel\psi\phi}{72\lambda\mu^3}\,,\\
    \phi'=&\frac{1}{72\mu^4}\left\{-16e^{\beta}\delta\lambda^2\mu\sin\gamma\left(w_\parallel-2 w_\perp+3\right)+\lambda^3\left[4\delta\left(w_\parallel-2w_\perp+3\right)-9\left(w_\parallel-2w_\perp+5\right)\right]\right.\\
    &\left.+36\lambda\mu^2\psi^2\left(w_\parallel-2w_\perp+1\right)+72\mu^3\psi\phi\left(w_\parallel-2 w_\perp-1\right)\right\}\,,\nonumber\\
    \lambda '=&\phi\,,\\
   \mu '=&\psi\,,
\end{align}
\end{subequations}
supplemented by the barotropic equation of state and auxiliary field equations
\begin{gather}\label{eq:densII}
    \nu=\beta\,,\quad \kappa\rho=\frac{\left(4\delta-9\right)\lambda^3-4\mu\left[4\delta\lambda^2 e^\beta\sin\gamma+9\mu\psi\left(2\mu\phi+\lambda\psi\right)\right]}{36\lambda\mu^4}\,.
\end{gather}

\subsection{Type III}\label{ssec:cosmofieldIII}
For type III, we largely follow the same procedure as for type II. First, we state the symmetric field equations:
\begin{subequations}\label{eq:feqsymIII}
\begin{align}
    E_{(00)}=&\frac{-16 \delta  \lambda ^2 \mu  \sinh\beta \sin\gamma+(2 \delta -9) \lambda ^3+4 \lambda  \mu ^2 \left[-8 \delta  \sin ^2\gamma +2 \delta  \mu ^2 (\gamma')^2+9(\mu')^2-9\right]+72 \mu ^3 \lambda' \mu'}{36 \lambda  \mu ^4}=\kappa\rho\,,\\
    E_{(11)}=&\frac{-8 \delta  \lambda^2  \mu  \sinh\beta \sin\gamma+(2 \delta -9) \lambda ^3+4\lambda \mu ^3 \left[2 \delta  \gamma' \left(\mu  \gamma'-2 \cosh\beta \sin\gamma \right)-9 \mu''\right]-36 \mu ^3 \left(\lambda' \mu'+\mu  \lambda''\right)}{36 \lambda  \mu ^2}=\kappa w_\perp\mu^2\rho\,,\\
    E_{(33)}=&\frac{\lambda ^2}{36 \mu ^4}\biggl\{32 \delta  \lambda  \mu  \sinh\beta \sin\gamma +3(9-2 \delta ) \lambda ^2\\
    &+4 \mu ^2 \left\{8 \delta  \sin ^2\gamma+2 \mu  \left[\delta  \gamma' \left(\mu  \gamma'-4 \cosh\beta \sin\gamma\right)-9 \mu''\right]-9\left[(\mu')^2+1\right]\right\}\biggl\}=\kappa w_\parallel\lambda^2\rho\,,\nonumber
\end{align}
\end{subequations}
and next the antisymmetric equations:
\begin{subequations}\label{eq:feqantisymIII}
\begin{align}
    E_{[14]}=&\frac{2\delta\lambda \sin\gamma\left(\lambda\cosh\beta-2\mu^2\sinh\beta\gamma'\right)}{9\mu^3}=0\,,\\
    E_{[23]}=&\frac{2 \delta  e^x}{9 \lambda  \mu }\biggl\{-\lambda ^2 \sinh \beta \cos\gamma+\lambda  \mu  \left\{\mu ^2 \gamma''-2 \left[\sin (2 \gamma )+\mu  \sinh\beta \sin\gamma \beta'+\mu' \left(\cosh\beta \sin\gamma-\mu  \gamma'\right)\right]\right\}\\
    &+\mu ^2 \lambda' \left(\mu  \gamma'-2 \cosh\beta \sin\gamma\right)\biggl\}=0\,.\nonumber
\end{align}
\end{subequations}
Let's begin with the antisymmetric equations. We see that $E_{[14]}=0$ has two solutions: $\gamma=N\pi,\ N\in\mathbb{Z}$ and
\begin{equation}\label{eq:gsolIII}
    \gamma'=\frac{\lambda\coth\beta}{2\mu^2}\,.
\end{equation}
Let's look at the first branch of solutions. By setting $\gamma=N\pi,\ N\in\mathbb{Z}$, the second antisymmetric equation becomes
\begin{equation}
    E_{[23]}\propto\frac{\delta e^x\lambda\sinh\beta}{\mu}=0\,,
\end{equation}
which we solve by setting $\beta=0$. The symmetric field equations can then be solved for $\rho$, $\lambda''$ and $\mu''$. Before doing so, we also introduce the auxiliary variables $\phi=\lambda'$ and $\psi=\mu'$ to get a first order system:
\begin{subequations}\label{eq:ressol1III}
\begin{align}
    \rho=&\frac{(2 \delta -9) \lambda ^3+36 \lambda  \mu ^2 \left(\psi ^2-1\right)+72 \mu ^3 \psi  \phi }{36 \kappa  \lambda  \mu ^4}\,,\\
    \phi'=&\frac{(2\delta -9) \lambda ^3 (w_{\parallel }-2 w_{\perp}+5)+36 \lambda  \mu ^2 \left(\psi ^2-1\right) (w_{\parallel }-2 w_{\perp}+1)+72 \mu ^3 \psi  \phi  (w_{\parallel}-2 w_{\perp}-1)}{72 \mu ^4}\,,\\
   \psi'=&-\frac{(2 \delta -9)\lambda ^2 (w_{\parallel }+3)}{72 \mu ^3}-\frac{w_{\parallel } \psi  \phi }{\lambda }-\frac{(w_{\parallel }+1) \left(\psi ^2-1\right)}{2 \mu }\,,\\
   \lambda '=&\phi\,,\\
   \mu '=&\psi\,.
\end{align}
\end{subequations}
We designate this branch as the ``restricted case'' and cover it in more detail in section \ref{ssec:restrictedsol}. In the other branch, we use eq.~\eqref{eq:gsolIII} to solve $E_{[23]}=0$ for $\beta'$:
\begin{equation}\label{eq:bsolIII}
    \beta'=-\frac{2 \left\{\mu  \coth \beta \left[\lambda  \left(2 \mu' -\lambda'  \operatorname{csch}\beta\csc \gamma \right)+2 \mu  \lambda' \right]+\lambda\cos \gamma\left(4 \mu \operatorname{csch}\beta+\lambda \csc \gamma\right)\right\}}{\lambda  \mu \left(\lambda  \operatorname{csch}^3\beta \csc \gamma+4 \mu \right)}\,.
\end{equation}
With this, we can solve the symmetric field equations again for $\rho$, $\lambda''$ and $\mu''$. After doing so and again substituting $\phi=\lambda'$ and $\psi=\mu'$, we get:
\begin{subequations}\label{eq:dynsysIII}
\begin{align}
    \gamma '=&\frac{\lambda \coth \beta}{2 \mu ^2}\,,\\
    \beta '=&-\frac{2 \left\{\mu  \coth \beta \left[\lambda  \left(2 \psi -\phi  \operatorname{csch}\beta\csc \gamma \right)+2 \mu  \phi \right]+\lambda\cos \gamma\left(4 \mu \operatorname{csch}\beta+\lambda \csc \gamma\right)\right\}}{\lambda  \mu \left(\lambda  \operatorname{csch}^3\beta \csc \gamma+4 \mu \right)}\,,\\
   \phi '=&\frac{1}{72 \mu ^4}\left\{-16 \delta  \lambda ^2 \mu \sinh \beta \sin \gamma\left(w_{\parallel }-2 w_{\perp}+3\right)\right.\\
   &+\lambda ^3 \left[2 \delta  \operatorname{csch}^2\beta \left(w_{\parallel }-2 w_{\perp}+1\right)+4 \delta  \left(w_{\parallel}-2 w_{\perp}+3\right)-9 \left(w_{\parallel }-2 w_{\perp}+5\right)\right]\nonumber\\
   &\left.+4 \lambda  \mu ^2 (w_{\parallel }-2 w_{\perp}+1) \left[9\psi ^2-\left(9+8 \delta  \sin^2\gamma\right)\right]+72 \mu ^3 \psi  \phi  (w_{\parallel }-2 w_{\perp}-1)\right\}\,,\nonumber\\
   \psi ^{\prime}=&-\frac{1}{72 \lambda  \mu ^3}\left\{16 \delta  \lambda ^2 \mu \sinh \beta \sin \gamma \left(\operatorname{csch}^2\beta-w_{\parallel}-1\right)\nonumber\right.\\
   &+\lambda ^3 \left[2 \delta  (w_{\parallel }-1)\operatorname{csch}^2\beta+4 \delta  (w_{\parallel }+1)-9 (w_{\parallel }+3)\right]\nonumber\\
   &\left.+4 \lambda  \mu ^2 (w_{\parallel }+1) \left[9 \psi ^2-\left(9+8 \delta  \sin ^2\gamma\right)\right]-72 \mu ^3 w_{\parallel } \psi  \phi \right\}\,,\nonumber\\
   \lambda '=&\phi\,,\\
   \mu '=&\psi\,,
\end{align}
\end{subequations}
with the additional equations
\begin{equation}\label{eq:densIII}
    \nu=\beta\,,\quad \kappa\rho=\frac{1}{72  \lambda  \mu ^4}\left\{144 \mu ^3 \psi  \phi -32 \delta  \lambda ^2 \mu \sinh \beta  \sin \gamma -\lambda ^3\left[9-2\delta\left(2+\operatorname{csch}^2\beta\right)\right]+8 \lambda  \mu ^2 \left[9 \psi ^2-\left(9+8 \delta  \sin^2\gamma\right)\right]\right\}\,.
\end{equation}
\subsection{Type IX}\label{ssec:cosmofieldIX}
Finally, we turn to type IX. The symmetric field equations are:
\begin{subequations}\label{eq:feqsymIX}
\begin{align}
    E_{(00)}=&\frac{16 \delta  \lambda ^2 \mu  \cosh \beta \sin \gamma+(2 \delta -9) \lambda ^3+4 \lambda  \mu ^2 \left[8 \delta  \sin ^2\gamma+2 \delta  \mu ^2 \left(\gamma '\right)^2+9 \left(\mu '\right)^2+9\right]+72 \mu ^3 \lambda ' \mu '}{36 \lambda  \mu ^4}=\kappa\rho\,,\\
    E_{(11)}=&\frac{8 \delta  \lambda^2  \mu  \cosh \beta \sin \gamma+(2 \delta -9) \lambda ^3+4\lambda \mu ^3 \left[2 \delta  \gamma ' \left(2 \sinh \beta \sin \gamma+\mu  \gamma
   '\right)-9 \mu ''\right]-36 \mu ^3 \left(\lambda ' \mu '+\mu  \lambda ''\right)}{36\lambda \mu ^2}=\kappa w_\perp\mu^2\rho\,,\\
    E_{(33)}=&-\frac{\lambda ^2}{36 \mu ^4}\left\{32 \delta  \lambda  \mu  \cosh \beta \sin \gamma+3 (2 \delta -9) \lambda ^2\right.\nonumber\\
    &\left.+4 \mu ^2 \left[8 \delta  \sin ^2\gamma-2 \mu  \left(\delta  \gamma' \left(4 \sinh \beta \sin \gamma+\mu  \gamma '\right)-9 \mu ''\right)+9 \left(\mu '\right)^2+9\right]\right\}=\kappa w_\parallel\lambda^2\rho\,,
\end{align}
\end{subequations}
and the antisymmetric equations are:
\begin{subequations}\label{eq:feqantisymIX}
\begin{align}
    E_{[14]}=&\frac{2\delta\lambda\sin\gamma\left(\lambda\sinh\beta-2\mu^2\cosh\beta\gamma'\right)}{9\mu^3}=0\,,\\
    E_{[23]}=&\frac{2\delta\sin x}{9\lambda\mu}\biggl\{\lambda^2\cos\gamma\cosh\beta+\mu^2\left(2\sin\gamma\sinh\beta+\mu\gamma'\right)\\
    &\left.+2\lambda\mu\left[\sin(2\gamma)+\mu\beta'\sin\gamma\cosh\beta+\mu'\left(\sin\gamma\sinh\beta+\mu\gamma'\right)+\frac{\mu^2\gamma''}{2}\right]\right\}=0\,.\nonumber
\end{align}
\end{subequations}
Now, $E_{[14]}=0$ has three branches of solutions: $(\gamma=N\pi,\, N\in\mathbb{Z})$, $(\gamma=\text{const.},\, \beta=0)$, and
\begin{equation}\label{eq:gsolIX}
    \gamma'=\frac{\lambda  \tanh \beta }{2 \mu ^2}\,.
\end{equation}
Let's look at each branch in order. In the first branch, $E_{[23]}$ becomes:
\begin{equation}
    E_{[23]}\propto\frac{\delta\lambda\cosh\beta\sin x}{\mu}\neq0\,,
\end{equation}
so this branch can be discarded. The second branch becomes:
\begin{equation}
    E_{[23]}\propto\frac{\delta\cos\gamma\sin x\left(\lambda+4\mu\sin\gamma\right)}{\mu}\,,
\end{equation}
and we see that this splits into two more branches, given by $\left(\gamma=\frac{2N+1}{2}\pi,\, N\in\mathbb{Z}\right)$ and $\lambda/\mu=-4\sin\gamma$. First, we solve the symmetric field equations in the former case:
\begin{subequations}\label{eq:ressol1}
\begin{align}
    \phi'=&\frac{1}{72 \mu ^4}\left\{16\delta\left(3+w_\parallel-2w_\perp\right)\lambda^2\mu\sgn\gamma-\left(9-2\delta\right)\left(5+w_\parallel-2w_\perp\right)\lambda^3\right.\\
   &\left.-72\left(1-w_\parallel+2w_\perp\right)\mu^3\psi\phi+4\left(1+w_\parallel-2w_\perp\right)\lambda\mu^2\left[8\delta+9\left(1+\psi^2\right)\right]\right\}\,,\nonumber\\
   \psi'=&\frac{1}{72 \lambda  \mu ^3}\left\{-16 \delta\left(2+w_\parallel\right)  \lambda ^2 \mu\sgn\gamma+\left(9-2\delta\right)\left(3+w_\parallel\right)\lambda^3\right.\\
   &\left.-72w_\parallel\mu^3\psi\phi-4\left(1+w_\parallel\right)\lambda\mu^2\left[8\delta+9\left(1+\psi^2\right)\right]\right\}\,,\nonumber\\
    \kappa\rho=&\frac{16 \delta  \lambda ^2 \mu\sgn\gamma-\left(9-2\delta\right)\lambda^3+72\mu^3\psi\phi+4\mu^2\lambda\left[8\delta+9\left(1+\psi^2\right)\right]}{36  \lambda  \mu ^4}\,,\\
    \lambda'=&\ \phi\,,\\
    \mu'=&\psi\,,
\end{align}
\end{subequations}
where we again introduced $\phi=\lambda'$ and $\psi=\mu'$. We will write the subsequent cases as first-order in the same way.\\
In the latter case, by substituting $\gamma=\text{const.}$, $\beta=0$ and $\lambda=-4\mu\sin\gamma$ into the symmetric field equations and solving for $\rho$ and $\psi'$, we get:
\begin{subequations}\label{eq:ressol2}
\begin{align}
    \psi'=&-\frac{1+w_\parallel-4(3+w_\parallel)\sin^2\gamma+(1+3w_\parallel)\psi^2}{2\mu}\,,\label{eq:ressol2a}\\
    \kappa\rho=&\frac{1-4\sin^2\gamma+3\psi^2}{\mu ^2}\,,\\
    0=&\sin\gamma\left[1+W-4(4+W)\sin^2\gamma+3W\psi^2\right]\,,\quad W=w_\parallel-w_\perp\,.\label{eq:ressol2c}
\end{align}
\end{subequations}
We designate the second main branch in its entirety as the ``restricted case'', similar to what was done for type III, and cover it in more detail in section \ref{ssec:restrictedsol}.

Next, we cover the third, and most general, main branch. First, we get an expression for $\beta'$:
\begin{equation}\label{eq:bsolIX}
    \beta '=-\frac{2 \left\{\lambda  \mu  \operatorname{sech}\beta \left[2 \sin \gamma \left(\psi  \sinh \beta+2 \cos\gamma\right)+\phi  \tanh \beta\right]+2 \mu ^2 \phi  \tanh\beta \sin\gamma+\lambda ^2 \cos\gamma\right\}}{\lambda  \mu  \left(\lambda  \operatorname{sech}^3\beta+4 \mu  \sin\gamma\right)}\,.
\end{equation}
Using eqs.~\eqref{eq:gsolIX} and~\eqref{eq:bsolIX}, we can solve the symmetric field equations for the third branch:
\begin{subequations}\label{eq:symsolIX}
\begin{align}
    \phi'=&\frac{1}{72 \mu ^4}\left\{16 \delta  \lambda ^2 \mu  \cosh \beta \sin\gamma \left(w_{\parallel }-2 w_{\perp}+3\right)\right.\\
   &+\lambda ^3 \left[-2 \delta \operatorname{sech}^2\beta \left(w_{\parallel }-2 w_{\perp}+1\right)+4 \delta  \left(w_{\parallel }-2 w_{\perp}+3\right)-9 \left(w_{\parallel }-2 w_{\perp}+5\right)\right]\nonumber\\
   &\left.+4  \mu^2\left\{ \lambda\left(w_{\parallel }-2 w_{\perp}+1\right) \left[8 \delta\sin^2\gamma +9\left (1+\psi^2\right)\right]+18 \mu \psi  \phi  \left(w_{\parallel }-2 w_{\perp}-1\right)\right\}\right\}\,,\nonumber\\
   \psi'=&\frac{1}{72 \lambda  \mu ^3}\left\{-16 \delta  \lambda ^2 \mu\operatorname{sech}\beta  \sin \gamma \left[w_{\parallel }+2+\left(w_{\parallel }+1\right) \sinh^2 \beta\right]\right.\nonumber\\
   &+\lambda ^3 \left[2 \delta  \left(w_{\parallel}-1\right) \operatorname{sech}^2\beta-4 \delta  \left(w_{\parallel }+1\right)+9 \left(w_{\parallel }+3\right)\right]\nonumber\\
   &\left.-4 \lambda  \mu ^2 \left(w_{\parallel }+1\right) \left[8 \delta  \sin ^2\gamma+9\left(1+\psi^2\right)\right]-72 \mu
   ^3 w_{\parallel } \psi  \phi \right\}\,,\nonumber\\
    \kappa\rho=&\frac{16 \delta  \lambda ^2 \mu  \cosh \beta \sin \gamma -\lambda ^3 \left[9-2 \delta\left(2-\operatorname{sech}^2\beta\right)\right]+4 \lambda  \mu ^2 \left\{8 \delta\sin^2\gamma +9\left[1+\psi^2\right]\right\}+72 \mu ^3 \psi  \phi }{36  \lambda  \mu ^4}\,.
\end{align}
\end{subequations}
Finally, we present this branch as a first-order system:
\begin{subequations}\label{eq:dynsysIX}
\begin{align}
    \gamma '=&\frac{\lambda  \tanh \beta }{2 \mu ^2}\,,\\
    \beta '=&-\frac{2 \left\{\lambda  \mu  \operatorname{sech}\beta \left[2 \sin \gamma \left(\psi  \sinh \beta+2 \cos\gamma\right)+\phi  \tanh \beta\right]+2 \mu ^2 \phi  \tanh\beta \sin\gamma+\lambda ^2 \cos\gamma\right\}}{\lambda  \mu  \left(\lambda  \operatorname{sech}^3\beta+4 \mu  \sin\gamma\right)}\,,\\
   \phi '=&\frac{1}{72 \mu ^4}\left\{16 \delta  \lambda ^2 \mu  \cosh \beta \sin\gamma \left(w_{\parallel }-2 w_{\perp}+3\right)\right.\\
   &+\lambda ^3 \left[-2 \delta \operatorname{sech}^2\beta \left(w_{\parallel }-2 w_{\perp}+1\right)+4 \delta  \left(w_{\parallel }-2 w_{\perp}+3\right)-9 \left(w_{\parallel }-2 w_{\perp}+5\right)\right]\nonumber\\
   &\left.+4  \mu ^2\left\{ \lambda\left(w_{\parallel }-2 w_{\perp}+1\right) \left[8 \delta\sin^2\gamma +9\left(1+\psi ^2\right)\right]+18 \mu \psi  \phi  \left(w_{\parallel }-2 w_{\perp}-1\right)\right\}\right\}\,,\nonumber\\
   \psi'=&\frac{1}{72 \lambda  \mu ^3}\left\{-16 \delta  \lambda ^2 \mu\operatorname{sech}\beta  \sin \gamma \left[w_{\parallel }+2+\left(w_{\parallel }+1\right) \sinh^2 \beta\right]\right.\nonumber\\
   &+\lambda ^3 \left[2 \delta  \left(w_{\parallel}-1\right) \operatorname{sech}^2\beta-4 \delta  \left(w_{\parallel }+1\right)+9 \left(w_{\parallel }+3\right)\right]\nonumber\\
   &\left.-4 \lambda  \mu ^2 \left(w_{\parallel }+1\right) \left[8 \delta  \sin ^2\gamma+9\left(1+\psi ^2\right)\right]-72 \mu
   ^3 w_{\parallel } \psi  \phi \right\}\,,\nonumber\\
   \lambda '=&\phi\,,\\
   \mu '=&\psi\,,
\end{align}
\end{subequations}
and
\begin{gather}\label{eq:densIX}
    \nu=\beta\,,\quad \kappa\rho=\frac{16 \delta  \lambda ^2 \mu  \cosh \beta \sin \gamma -\lambda ^3 \left[9-2 \delta\left(2-\operatorname{sech}^2\beta\right)\right]+4 \lambda  \mu ^2 \left[8 \delta\sin^2\gamma +9\left(1+\psi ^2\right)\right]+72 \mu ^3 \psi  \phi }{36  \lambda  \mu ^4}\,.
\end{gather}
We will investigate the properties of this system in more detail in section \ref{ssec:gensol}.

\section{Solutions}\label{sec:solution}
With the full dynamical systems now at hand, we can shift our focus to specific cases of interest, beginning with a look at the previously mentioned restricted cases in Bianchi III and IX in section~\ref{ssec:restrictedsol}, followed by the general systems in section~\ref{ssec:gensol}. Note that an investigation of the full solution space of the presented dynamical systems would exceed the scope of this article, and so we restrict ourselves to a few example solutions and their physical interpretation. In particular, we show a few solutions in which the ratio \(\lambda/\mu\) of the scale factors remains constant, thus resembling an isotropic expansion; we call these projective fixed points, as they can be understood as fixed points of a projective space obtained from the dynamical variables.

\subsection{Restricted solutions}\label{ssec:restrictedsol}
First, we look at the type III case given by eq.~\eqref{eq:ressol1III}, where $\left(\gamma=N\pi,\, N\in\mathbb{Z}\right)$ and $\beta=0$. We note that it is possible to restate the expression for $\rho$ as:
\begin{equation}\label{eq:resdens}
    \frac{\phi'}{\lambda}+2\frac{\psi'}{\mu}=-\frac{1+w_\parallel+2w_\perp}{2}\kappa\rho\,,
\end{equation}
which is analogous to one of the Friedmann equations without a cosmological constant and states that the existence of non-exotic matter causes the universe to experience decelerating expansion. In addition, we can rewrite the equations for $\phi'$ and $\psi'$ as:
\begin{subequations}\label{eq:resscales}
\begin{align}
    \frac{\phi'}{\lambda}=&-\frac{1-w_{\parallel }+2 w_{\perp}}{2} \kappa \rho-k(t)+\frac{\psi ^2}{\mu ^2}\,,\\
    \frac{\psi'}{\mu}=&-\frac{w_\parallel}{2} \kappa \rho +\frac{1}{2}\left(k(t)-\frac{\psi ^2}{\mu ^2}\right)\,,\\
    k(t)=&\frac{(9-2\delta)\lambda^2+12\mu^2}{12\mu^4}\,.
\end{align}
\end{subequations}
We can also extract some interesting behaviour with the transformation $\lambda=R\mu$, where $R=R(t)$. By applying this to eq.~\eqref{eq:ressol1III}, we find the following equation for $R''$:
\begin{equation}
    R''=\frac{R \left[(2 \delta -9) R^2-9\right]}{9 \mu ^2}+R'\frac{ \psi }{\mu } \left(2W-3\right)+\frac{WR\left[(2 \delta -9) R^2+36 \left(3 \psi ^2-1\right)\right]}{36 \mu ^2}\,,
\end{equation}
where $W=w_\parallel-w_\perp$. We see that the only solution to this equation that is independent of $\psi$ is $R=\pm\frac{3}{\sqrt{2\delta-9}}$ with $W=0$. To ensure that both scale factors remain real-valued, we must then impose $\delta>9/2$. With this, the remaining equations become, after some rearranging and rescaling $\mu\rightarrow\mu/2$:
\begin{equation}
    \frac{\kappa\rho}{3}=\left(\frac{\mu'}{\mu}\right)^2-\frac{1}{\mu^2}\,,\quad \frac{\mu''}{\mu}=-\frac{\kappa\rho}{6}\left(1+3w\right)\,,
\end{equation}
where $w_\parallel=w_\perp=w$. We see that these are the Friedmann equations with no cosmological constant and negative curvature. To investigate the stability of this system, we calculate its Jacobian for $R'$ and $R$:
\begin{equation}\label{eq:jacobian_III}
    \left.J^j_i\right|_{F=F_0}=\left.\frac{\partial F'_i}{\partial F_j}\right|_{F=F_0}=\frac{1}{\mu}\begin{pmatrix}
        -3\psi & \frac{2}{\mu} \\
        1 & 0
    \end{pmatrix},
\end{equation}
where  $F_i=(R'\,,R)$ and $F_0$ denotes the fixed point in phase space. This matrix has two eigenvalues $\Lambda_i$:
\begin{equation}
    \Lambda_{1,2}=-\frac{3\psi\pm\sqrt{9\psi^2+8}}{2\mu}\,.
\end{equation}
One of these is always positive for any real $\psi$ and $\mu$, so this system is unstable. In addition, it was shown in~\cite{Bahamonde:2024zkb} that 1PNGR is free of instabilities only when $\delta\leq0$. As this solution requires $\delta>9/2$, we may discard it as pathological.

Next, we look at the type IX case given by eq.~\eqref{eq:ressol1}, where $\left(\gamma=\frac{2N+1}{2}\pi,\, N\in\mathbb{Z}\right)$ and $\beta=0$. Here, we find that we can again rearrange the field equations to get eq.~\eqref{eq:resdens}. As such, the effect of matter on the expansion of the universe is the same. The equations for $\phi'$ and $\psi'$ can also be written in the same form as~\eqref{eq:resscales}, but with a different $k(t)$:
\begin{equation}
    k(t)=\frac{3(9-2\delta)\lambda^2-4\mu[(9+8\delta)\mu+8\delta\lambda\sgn\gamma]}{36\mu^4}\,.
\end{equation}
We can also perform a similar analysis to the type III restricted case, but we will see in the next section that we can use the general type IX system to obtain the same results, so we will not go through it here. Instead, we look at the second type IX restricted case given by eq.~\eqref{eq:ressol2}. First, we must solve eq.~\eqref{eq:ressol2c}, for which we have two options: keep $W\neq0$ and solve for some $\psi$, or set $W=0$ and solve for $\gamma$. In the first case, we can solve eq.~\eqref{eq:ressol2c} to get $\psi=\text{const.}$ Using this, we can set $\psi'=0$ and solve eqs.~\eqref{eq:ressol2a} and \eqref{eq:ressol2c} simultaneously for $\sin\gamma$ and $\psi$. However, we find that no real solutions exist, so we can discard this case and set $W=0$. This allows us to solve eq.~\eqref{eq:ressol2c} for $\sin\gamma=\pm1/4$ instead\footnote{The sign choice determines whether $\lambda$ and $\mu$ have the same or opposite signs.}, which we can then substitute into the other field equations to get:
\begin{equation}
    \psi'=-\frac{\left(1+3w\right)\left(1+4\psi^2\right)}{8\mu}\,,\quad\kappa\rho=\frac{3\left(1+4\psi^2\right)}{4\mu^2}\,,\quad w=w_\parallel=w_\perp.
\end{equation}
Finally, we can rescale the scale factor as $\mu\rightarrow\mu/2$ and rearrange these equations into a more familiar second-order form:
\begin{equation}\label{eq:feqfriedmannIX}
    \frac{\kappa\rho}{3}=\left(\frac{\mu'}{\mu}\right)^2+\frac{1}{\mu^2}\,,\quad \frac{\mu''}{\mu}=-\frac{\kappa\rho}{6}\left(1+3w\right)\,,
\end{equation}
which we recognise as the Friedmann equations without a cosmological constant and with positive curvature.

\subsection{General solutions}\label{ssec:gensol}
It is clear that the general dynamical systems \eqref{eq:dynsysII}, \eqref{eq:dynsysIII}, and \eqref{eq:dynsysIX} are difficult to analyse without resorting to numerical methods or simplifying the systems. For now, we will focus on the latter method. As we wish to keep the scale factors dynamical, a natural simplification to consider is one in which $\gamma'=\beta'=0$. From the dynamical systems, it is immediately clear that the type II and III systems do not support such a configuration, as for both $\gamma'\neq0$ and $\beta'\neq0$ for any real constant value of $\gamma$ and $\beta$. This is not the case for the type IX system. In fact, we find that we can apply the parameter choices from the first type IX restricted case to eq.~\eqref{eq:dynsysIX} to recover eq.~\eqref{eq:ressol1}. This is unlike the type III and the second type IX restricted cases, which are not subcases of the respective general ones. To investigate this system in detail, we first apply the transformation $\lambda= -R \mu\sgn\gamma$\footnote{The $\sgn\gamma$ factor could also be absorbed into $R$, but we use it here to simplify the following equations.} and set $\gamma=\text{const.}$, $\beta=0$. With this, we find that the equation for $\beta'$ from eq.~\eqref{eq:dynsysIX} becomes $0=\beta'\propto\cos\gamma$, so we must also apply $\left(\gamma=\frac{2N+1}{2}\pi,\, N\in\mathbb{Z}\right)$\footnote{For brevity, we shorten this to $\gamma=\pm\pi/2$ without loss of generality.}. We then get the following equation for $R''$:
\begin{equation}\label{eq:dynsysIX_R}
    R''=\frac{R(1-R)\left[R(9-2\delta)+9+8\delta\right]}{9 \mu ^2}+R'\left(2W-3\right)\frac{\psi}{\mu}+\frac{R W \left[4-R^2+\frac{2}{9} \delta  (4-R)^2+12 \psi ^2\right]}{4 \mu ^2}\,,
\end{equation}
where $W=w_\parallel-w_\perp$. Note that we can remove all terms proportional to $\psi$ by setting $R'=W=0$. As this implies that $R=\text{const.}=R_0$, then the numerator of the first term also becomes a constant. We can then solve $R''=0$ for $R_0$ to get:
\begin{equation}\label{eq:solR_fixed}
    R_0=1\quad \text{or}\quad R_0=-\frac{9+8\delta}{9-2\delta}\,,\quad \text{assuming } R_0\neq0\,.
\end{equation}
We now have a projective fixed point for the functions $\gamma$, $\beta$ and $R$ at $\gamma=\pm\pi/2$, $\beta=0$ and eq.~\eqref{eq:solR_fixed}. Our only remaining dynamical variables are $\rho$, $\psi$ and $\mu$, governed by the following equations:
\begin{equation}
    \kappa\rho=\frac{4-R_0^2+\frac{2}{9}\delta\left(4-R_0\right)^2+12 \psi^2}{36\mu^2}\,,\quad \psi'=-\frac{(1+3w)\left[(4-R_0^2+\frac{2}{9}\delta(4-R_0)^2+12\psi^2\right]}{24\mu}\,,\quad \mu'=\psi\,,
\end{equation}
where $w_\parallel=w_\perp=w\geq0$. Of additional interest is the stability of this configuration. To determine this, we calculate its Jacobian for $\gamma$, $\beta$, $R'$ and $R$ at the fixed point (labelled as $F_0$):
\begin{equation}\label{eq:jacobian_IX}
    \left.J^j_i\right|_{F=F_0}=\left.\frac{\partial F'_i}{\partial F_j}\right|_{F=F_0}=\frac{1}{\mu}\begin{pmatrix}
        0 & -\frac{R_0\sgn\gamma}{2} & 0 & 0 \\
        2\sgn\gamma & -2\psi & 0 & 0 \\
        0 & 0 & -3\psi & \frac{9+8\delta-20\delta R_0-3(9-2\delta)R_0^2}{9\mu} \\
        0 & 0 & 1 & 0
		\end{pmatrix}\,,
\end{equation}
where \(F_i=(\gamma, \beta, R', R)\). Next, we find its eigenvalues $\Lambda_i$. The result is:
\begin{equation}
    \Lambda_{1,2}=\frac{-\psi\pm\sqrt{\psi^2-R_0}}{\mu}\,,\quad \Lambda_{3,4}=\frac{-9\psi\pm\sqrt{81\psi^2+4\left[9+8\delta-20\delta R_0-3(9-2\delta)R_0^2\right]}}{6\mu}\,.
\end{equation}
These must all have a negative real part to ensure stability, which holds for $\Lambda_{1,2}$ when $\psi/\mu>0$ and $R_0>0$. Note that the latter condition can be satisfied with any one of the following:
\begin{equation}
    R_0=1\,,\quad \delta<-\frac{9}{8}\,,\quad \delta>\frac{9}{2}\,.
\end{equation}
As noted above, any instance of 1PNGR with $\delta>0$ is unstable. In addition, results from~\cite{Asukula:2023akj} show that some static and spherically symmetric spacetimes in 1PNGR require $\delta>-\frac{9}{8}$ to preserve the Lorentzian signature of the metric.
Therefore, the best option is to choose $R_0=1$. This somewhat simplifies the eigenvalues:
\begin{equation}\label{eq:jacobian_eigvalIX}
    \Lambda_{1,2}=\frac{-\psi\pm\sqrt{\psi^2-1}}{\mu}\,,\quad \Lambda_{3,4}=\frac{-9\psi\pm\sqrt{81\psi^2-24(3+\delta)}}{6\mu}\,.
\end{equation}
We now see that, if $\psi/\mu>0$, then $\Lambda_{3,4}$ also have negative real parts and the system is stable.

Another topic of interest is the matter of perturbations around this point. Using the Jacobian derived above, we can construct the perturbations of $\gamma$, $\beta$ and $R$ (denoted by a subscript $p$) around the projective fixed point at $\gamma_0=\pm\pi/2$, $\beta_0=R'=0$ and $R_0=1$:
\begin{equation}
    \gamma_p'=-\frac{\sgn\gamma_0}{2\mu}\beta_p\,,\quad \beta_p'=\frac{2}{\mu}\left(\gamma_p\sgn\gamma_0-\psi\beta_p\right)\,,\quad R_p''=-\frac{2(3+\delta)}{3\mu^2}R_p\,.
\end{equation}
These equations are not analytically solvable, as we do not know the exact expression for $\mu$. We see, though, that the term in front of $R_p$ in the third equation is always negative if $\delta>-3$, which implies that $R_p$ is a periodic function in this case. The remaining equations are coupled and also contain $\psi$, which makes their analysis unfeasible.

Before concluding, we briefly turn back to the general type II system and apply the transformation $\lambda=R\mu$. As before, we find that any dependence on $\psi$ can be eliminated by choosing $R'=w_\parallel-w_\perp=0$. The resulting equation for $R''$ is:
\begin{equation}\label{eq:Rdd_II}
    R''=\frac{R^2\left[(2\delta-9)R-8\delta e^\beta\sin\gamma\right]}{9\mu^2}\,.
\end{equation}
We see that we have $R''=0$ if $R=\frac{8\delta e^\beta\sin\gamma}{2\delta-9}$, but in this case we also require that $e^\beta\sin\gamma=\text{const.}$ We can calculate the derivative of this expression and substitute in $\gamma'$ and $\beta'$ from eq.~\eqref{eq:dynsysII} alongside $\lambda=R\mu$ and $R'=w_\parallel-w_\perp=0$ to get:
\begin{equation}
    \frac{d}{dt}\left(e^\beta\sin\gamma\right)=\frac{\left(R-4e^\beta\sin\gamma\right)\psi}{2\mu}\,,
\end{equation}
which is 0 when $R=4e^\beta\sin\gamma$ and implies that this and eq.~\eqref{eq:Rdd_II} are simultaneously 0 only when $\frac{2\delta}{2\delta-9}=1$, which has no solutions. Therefore, projective fixed points of this kind are completely prohibited in the Bianchi type II system.

\section{Conclusion}\label{sec:conclusion}
In this article, we have used the most general teleparallel geometries corresponding to the Bianchi types II, III and IX to construct cosmological solutions to the 1PNGR field equations. We find that the tetrads associated with these geometries possess additional symmetries that allow us to remove two of the six tetrad functions. In addition, the type III and IX tetrads possess a constant parameter $n$, which is also eliminated with certain transformations of the coordinates and tetrad functions. After deriving the most general energy-momentum tensors for the same Bianchi types, we derived the field equations and expressed them as a dynamical system. For these dynamical systems, we have found two types of solutions: the general solutions, in which all tetrad functions are dynamical, and the so-called restricted solutions, in which two tetrad functions are non-dynamical. We find that the non-dynamical functions in the restricted cases are always $\gamma$ and $\beta$, which do not appear in the metric, leaving the scale factor functions $\lambda$ and $\mu$ dynamical.

In addition, we find that the restricted cases only appear for the type III and IX geometries. For the type III restricted case, we find that the transformation $\lambda=R(t)\mu$ reveals an unstable projective fixed point at $R(t)=\pm\frac{3}{\sqrt{2\delta-9}}$, where $\delta$ is the 1PNGR parameter. However, this solution then also requires a positive $\delta$, which makes the theory itself unstable. The type IX cases are more interesting, producing two different branches. One of these reproduces the Friedmann equations with $\Lambda=0$ and positive curvature, albeit using a different metric than the usual FLRW. The second branch, with a similar transformation to the type III case, reveals another projective fixed point at $R(t)=1$. In this case, it is possible to keep both the fixed point and the wider theory stable by imposing $\mu'/\mu>0$ and also possible to investigate perturbations of $R(t)$ around this point, which reveal an oscillating behaviour. Finally, in the type II case, we find that neither a restricted case nor any fixed point is permitted.

The work we presented here invites numerous directions for further research. The most obvious line of research is to continue investigating Bianchi type II, III, and IX solutions of 1PNGR, and, in particular, to match these solutions with observations of the currently known evolution of the Universe. Due to the larger number of dynamical functions compared to the well-known homogeneous and isotropic cosmology, and therefore a larger space of solutions, this is a tedious task, which we aim to address with both analytic and numerical methods and computer support. Of particular interest is the question of whether a combination of early-time inflation and late-time accelerating expansion can be achieved without invoking dark energy. We also note that all of the results derived in sections~\ref{sec:geometry} and~\ref{sec:enmom} are theory-agnostic, which means they can also be used to derive the most general Bianchi type II, III and IX field equations of other metric teleparallel gravity theories, such as \(f(T)\) gravity or scalar-tensor theories, and subsequently study their solutions.

Besides these background solutions, another line of future research is to study perturbations around the Bianchi cosmological tetrads and energy-momentum tensors discussed here, derive their dynamical equations for the aforementioned classes of metric teleparallel gravity theories, and investigate whether these anisotropic backgrounds offer a potential solution to the strong coupling problem in cosmology.

%

\begin{acknowledgments}
The authors gratefully acknowledge the full financial support by the Estonian Research Council through the Personal Research Funding project PRG2608 and the Center of Excellence TK202 ``Fundamental Universe''.
\end{acknowledgments}

\bibliography{references}

\end{document}